\documentclass[12pt]{article}

\usepackage{graphicx}
\usepackage{amsmath}
\usepackage{amssymb}
\usepackage{hyperref}
\usepackage{times}
\usepackage{fullpage}
\usepackage{ifthen}
\usepackage{deluxetable}
\usepackage{placeins}

\newcommand{\physrep}{{Phys. Rep.}}
\newcommand{\aapr}{{Astron. Astrophys. Rev.}}
\newcommand{\mnras}{{Mon. Not. R. Astron. Soc.}}
\newcommand{\apj}{{Astrophys. J.}}

\newcommand{\aap}{{Astron. Astrophys.}}

\newcommand{\aj}{{Astron. J.}}
\newcommand{\apjl}{{Astrophys. J.}}
\newcommand{\pasj}{{Publ. Astron. Soc. Jpn.}}

\newcommand{\ssr}{{Space Sci. Rev.}}

\newcommand{\RNum}[1]{\uppercase\expandafter{\romannumeral #1\relax}}

\newenvironment{addendum}{%
    \setlength{\parindent}{0in}%
    \small%
    \begin{list}{Acknowledgements}{%
        \setlength{\leftmargin}{0in}%
        \setlength{\listparindent}{0in}%
        \setlength{\labelsep}{0em}%
        \setlength{\labelwidth}{0in}%
        \setlength{\itemsep}{12pt}%
        }
    }
    {\end{list}\normalsize}

\newcommand*{\addendumlabel}[1]{\textbf{#1}\hspace{1em}}

\newenvironment{methods}{%
    \section*{Methods}%
    \setlength{\parskip}{6pt}%
    }{}

\title{Observations of a Pre-Merger Shock in Colliding Clusters of Galaxies}

\author{Liyi Gu$^{1,2}$, Hiroki Akamatsu$^{2}$, Timothy W. Shimwell$^{3,4}$, Huib T. Intema$^{4,5}$, \\ 
Reinout J. van Weeren$^{4}$, 
Francesco de Gasperin$^{6,4}$, Fran\c{c}ois Mernier$^{7,8,2}$, \\ 
Junjie Mao$^{9,2}$, Igone Urdampilleta$^{2,4}$, Jelle de Plaa$^{2}$, \\
Viral Parekh$^{10, 11}$, Huub J. A. R{\"o}ttgering$^{4}$, \& Jelle S. Kaastra$^{2,4}$ }

\begin{document}

\maketitle

\newenvironment{affiliations}{%
    \setcounter{enumi}{1}%
    \setlength{\parindent}{0in}%
    \slshape\sloppy%
    \begin{list}{\upshape$^{\arabic{enumi}}$}{%
        \usecounter{enumi}%
        \setlength{\leftmargin}{0in}%
        \setlength{\topsep}{0in}%
        \setlength{\labelsep}{0in}%
        \setlength{\labelwidth}{0in}%
        \setlength{\listparindent}{0in}%
        \setlength{\itemsep}{0ex}%
        \setlength{\parsep}{0in}%
        }
    }{\end{list}\par\vspace{12pt}}

\begin{affiliations}
 \item RIKEN High Energy Astrophysics Laboratory, 2-1 Hirosawa, Wako, Saitama 351-0198, Japan
 \item SRON Netherlands Institute for Space Research, Sorbonnelaan 2, 3584 CA Utrecht, the Netherlands
 \item ASTRON, the Netherlands Institute for Radio Astronomy, Postbus 2, 7990 AA, Dwingeloo, The Netherlands
 \item Leiden Observatory, Leiden University, PO Box 9513, NL-2300 RA Leiden, The Netherlands
 \item International Centre for Radio Astronomy Research, Curtin University, GPO Box U1987, Perth, WA 6845, Australia
 \item Hamburger Sternwarte, Universit{\"a}t Hamburg, Go jenbergsweg 112, 21029, Hamburg, Germany
 \item MTA-E{\"o}tv{\"o}s University Lend{\"u}let Hot Universe Research Group, P{\'a}zm{\'a}ny P{\'e}ter 
s{\'e}t{\'a}ny 1/A, Budapest, 1117, Hungary
 \item Institute of Physics, E{\"o}tv{\"o}s University, P{\'a}zm{\'a}ny P{\'e}ter s{\'e}t{\'a}ny 1/A, Budapest, 1117, Hungary
 \item Department of Physics, University of Strathclyde, Glasgow, G4 0NG, UK
 \item Raman Research Institute, C. V. Raman Avenue, Sadashivanagar, Bangalore 560080, India
 \item Department of Physics \& Electronics, Rhodes University, PO Box 94, Grahamstown, 6140, South Africa
\end{affiliations}

\renewenvironment{abstract}{%
    \setlength{\parindent}{0in}%
    \setlength{\parskip}{0in}%
    \bfseries%
    }{\par\vspace{0pt}}
\begin{abstract}

Clusters of galaxies are the largest known gravitationally-bound structures in the Universe.
When clusters collide, they create merger shocks on cosmological scales, which transform
most of the kinetic energy carried by the cluster gaseous halos into heat \cite{ryu2003, bykov2008, bykov2019}.
Observations
of merger shocks provide key information of the merger dynamics, and enable insights 
into the formation and thermal history of the large-scale structures. Nearly all of the merger shocks
are found in systems where the clusters have already collided \cite{maxim2007, russell2010, akamatsu2013, ogrean2013, das2016, igone2018, feretti2012, vw2017, vw2019}, knowledge of shocks in the pre-merger phase is a crucial missing ingredient \cite{maxim1999, kato2015}.
Here we report on the discovery of a unique shock in a cluster pair 1E~2216 and 1E~2215. The two clusters
are observed at an early phase of major merger.
Contrary to all the known merger shocks observed ubiquitously on 
merger axes, the new shock propagates outward along the equatorial plane of the merger. 
This discovery uncovers an important epoch in the formation of massive clusters, when the rapid approach
of the cluster pair leads to strong compression of gas along the merger axis. Current theoretical models
\cite{akahori2010, ha2018} predict that the bulk of the shock energy might be 
dissipated outside the clusters, and eventually turn into heat of the pristine gas in the circum-cluster 
space.

\end{abstract}

Here we present a study of an pre-merger cluster pair 1E~2216.0-0401 and 1E~2215.7-0404 
at a redshift of 0.09, primarily with 149/139/44 kiloseconds of 
{\it Chandra}/{\it XMM-Newton}/{\it Suzaku} X-ray observations. As shown in Figure~1, the 
cores of the two clusters are separated by a distance of 640~kpc. 
The central regions of the two clusters have similar ICM temperatures ($4.7 \pm 0.1$~keV and $5.4 \pm 0.1$~keV), suggesting a nearly equal-mass merger.
Most interestingly, the {\it Chandra} image reveals a wedge-like feature extended towards the southeast on the equatorial plane between the two clusters (Figure~2). The wedge is clearly seen in the X-ray residual map where the X-ray
haloes of the two clusters are modeled (Supplementary Figure~1) and subtracted, suggesting that it 
cannot just be the overlapping of the two clusters. The leading edge of the wedge is located about 450~kpc away from 
the collision axis. A radial profile of X-ray surface
brightness as a function of distance from the collision axis shows a clear 
discontinuity across the edge (Supplementary Figures~2-4), 
indicating a sharp drop in gas density. The surface brightness
steepening at the discontinuity significantly (with 98\% confidence) deviates from a smooth King profile or an exponential profile. 


The equatorial plane is found to be hotter than the surrounding area in the projected temperature map
made with the {\it XMM-Newton} data (Figure~3 and Supplementary Figure~6). The peak temperature, $\sim 8.1 \pm 1.4$~keV, 
is found at the leading edge of the wedge. A secondary peak is seen at the north side of the equatorial plane.
No hot structure
is seen at the other directions around the two clusters, suggesting that the observed temperature increase cannot be merely a 
temperature gradient internal to each of the clusters. Combining the temperature and surface brightness
distribution around the wedge, we find a significant pressure discontinuity at its leading edge. This strongly 
favours the presence
of a shock, rather than a cold front, at the location of the observed surface brightness discontinuity. An X-ray 
entropy map 
reveals the signs of heating extending towards a larger distance than the observed shock front, 
hinting for previous equatorial shocks at an earlier stage of the merger (Supplementary Figure~7).

We use different approaches to calculate the velocity of the observed shock. By measuring the surface brightness
discontinuity, the ICM behind the shock front is found to be compressed by a mean value of $2.1 \pm 0.5$, giving a shock 
Mach number $\mathcal{M} = 1.8 \pm 0.5$. 
A better way to estimate the shock speed is to determine the temperature jump 
at the shock (Supplementary Figure~5),
because this method is less affected by the unknown shock geometry \cite{akamatsu2015, akamatsu2016, hallman2018, storm2018}.
The average temperature jump obtained from the different 
instruments weighted by their errors is $1.6 \pm 0.4$, giving a shock Mach number $\mathcal{M} = 1.6 \pm 0.3$. The observed temperature 
jump is nearly consistent with the value expected from adiabatic compression, suggesting that the ICM heating is 
mostly done with the shock compression. Assuming that the shock is propagating from the collision axis with an increasing velocity \cite{ha2018}, the shock has an age $> 250$~Myr. Comparing it to the age of the possible axial shock
($\sim 50-100$~Myr) reported in \cite{akamatsu2016}, the equatorial shock seems to have formed substantially earlier. 

Numerical simulations \cite{akahori2010,ha2018} predict that merger shocks along the equatorial plane should appear in pairs, however
in our target only one is visible in the X-ray image (Figure~2). 
We then searched for signatures of shocks using 144~MHz radio observations from the LOw Frequency ARray (LOFAR), 
and 235~MHz and 325~MHz observations from the Giant Metrewave Radio Telescope (GMRT). 
The radio images reveal a complex
radio structure at the northwest of the equatorial plane, where a second equatorial shock is expected. The radio feature is 
mainly dominated by a pair of radio tails of $\sim 330$~kpc size (source A in Figure~2), and a bright patch 
with a narrow extension of $\sim 450$~kpc towards northwest (source B). A close inspection
of the LOFAR and GMRT images shows that the radio tails are probably associated with
a reported cluster member galaxy (Supplementary Figure~8; \cite{jones2009}). 
The 610~MHz GMRT image reveals 
a compact radio core in the probable host galaxy. Therefore, we conclude 
that source A is associated with a member AGN, but the nature of source B still remains unclear.

The spectral measurement between 144~MHz and 325~MHz reveals a steep spectrum of the radio tails (source A), with an average
spectral index of $-2.3 \pm 0.1$. Naturally, the radio spectrum is expected to steepen away from the AGN due to ageing effect, 
however, the observed spectral index distribution (Figure~3 and Supplementary Figure~9) appears to be too flat to be explained
by radio radiation. The cooling cosmic ray particles must have been re-energized, probably through a Fermi-type shock
acceleration process \cite{ensslin2001, ensslin2002}.
The re-acceleration balances out the radiative loss, inducing a 
flattening of the spectral index at the shock \cite{vw2017}. 
Furthermore, source A seems to overlap, in projection, 
with the heated ICM structure 
($k$T $\geq 7.0$~keV) at the northwest of the equatorial plane (Figure~3). It is therefore likely that the re-acceleration of the radio tails and 
the ICM heating
are related, probably powered by the same shock (a twin to the southeast shock) or an adiabatic compression triggered by the merger. 

Contrary to the north region, we do not detect significant radio emission at the southeast shock. 
If the north radio sources are revived from cooling cosmic ray particles, we may conclude that the bulk of
such particles is not present at the southeast. The direct injection of relativistic particles out of the
thermal pool might not be efficient, as the observed shock speed, $\mathcal{M} \sim 1.6$, is 
smaller than minimal Mach number required for acceleration in several theoretical models
\cite{hoeft2011, pinzke2013, vink2014}.

The X-ray observations of the sharp surface brightness discontinuity and the associated high temperature structure 
provide the best evidence to date for the presence of an equatorial shock in an early stage of on-axis merger.  The powerful shock at southeast 
contains a kinetic energy flow rate of $\sim 8.1 \times 10^{44}$ erg s$^{-1}$, about two times of the combined X-ray luminosity of the two clusters. The northwest shock, if exists, carries a similar amount of kinetic
energy as the southeast one.

Our observation resembles the results from the state-of-the-art theoretical models. The latest hydrodynamic codes \cite{akahori2010, ha2018} predicted that the very first pair of merger shocks 
should occur before the core collision, and propagate outward along the equatorial plane. These shocks are 
formed by the interaction of the outer regions of the two clusters. Previously, there is 
only tentative observational evidence for this theory, including the X-ray temperature increase between the early-stage mergers A85 \cite{schenck2014},
A115 \cite{hallman2018}, A399-A401 \cite{akamatsu2017}, and A1758 \cite{botteon2018},
 as well as a Sunyaev-Zel'dolvich (SZ) decrement reported by \cite{rumsey2017} between
the two sub-clusters of CIZA~J2242+5301. 
However, A85 and A115 are off-axis mergers, the high temperature regions could be explained alternatively by 
a side-by-side interaction of regular shocks driven by the infalling subclusters.
For the on-axis mergers A399-A401 and A1758, no sharp X-ray surface brightness discontinuity could be found with the available data.
Hence, it is unclear whether the observed temperature breaks are caused by shocks, particularly equatorial shocks, or by a milder adiabatic compression of the 
two merger clumps during the early merger stage. 
Similarly, adiabatic compression could also explain the SZ feature in CIZA~J2242+5301. Moreover,
CIZA~J2242+5301 is already at a late stage of merger. 
It is hard to detect a clean feature from its early stage as it would have already been mixed.
We believe that the detection of the equatorial shock in 1E~2216.0-0401 and 1E~2215.7-0404 provides the ideal and most
direct support for the theoretical model.

The discovery of equatorial shock completes the inventory of merger shocks. It reveals 
an important merging epoch right before the core collision, when shocks are formed at the contact 
interface and propagate vertically to the merger axis. 
For the first time, we measure the energy release by the equatorial shock,
which may prove to be one of the important dissipation processes of major mergers. As shown in 
the numerical simulations \cite{akahori2010,ha2018}, the equatorial shocks have the highest velocity among
merger shocks, allowing them to reach large distance
($\geq 3$ Mpc) before they disappear. These shocks, together with some axial shocks that form later, 
might dump 
a significant portion of kinetic energy beyond the cluster boundary, providing a feasible way
to heat the inter/circum-galactic medium in the cluster surroundings.


\begin{addendum}
\item[Correspondence] Correspondence and requests for materials
should be addressed to L.G.~(email: liyi.gu@riken.jp).

\item[Acknowledgements] L.G. is supported by the RIKEN Special Postdoctoral Researcher Program. 
SRON is supported financially by NWO, the Netherlands Organization for Scientific Research. 
R.J.vW acknowledges support from the VIDI research programme with project number 639.042.729, which is financed by the Netherlands Organisation for Scientific Research (NWO). 
F.M. is supported by the Lend{\"u}let LP2016-11 grant awarded by the Hungarian Academy of Sciences. J.M. acknowledges support by STFC (UK) through the UK APAP network grant ST/R000743/1.

This research has made use of data obtained from the Chandra Data Archive and the Chandra Source Catalog, and software provided by the Chandra X-ray Center (CXC) in the application packages CIAO, ChIPS, and Sherpa. This research is partly based on observations obtained with XMM-Newton, an ESA science mission with instruments and contributions directly funded by ESA Member States and NASA. We thank the staff of the GMRT that made these observations possible. GMRT is run by the National Centre for Radio Astrophysics of the Tata Institute of Fundamental Research. This paper is based in part on data obtained with the International LOFAR Telescope (ILT) under project code LC7\_003 and LC9\_014. LOFAR is the Low Frequency Array designed and constructed by ASTRON. It has observing, data processing, and data storage facilities in several countries, that are owned by various parties (each with their own funding sources), and that are collectively operated by the ILT foundation under a joint scientific policy. The ILT resources have benefited from the following recent major funding sources: CNRS-INSU, Observatoire de Paris and Universit{\'e} d'Orl{\'e}ans, France; BMBF, MIWF-NRW, MPG, Germany; Science Foundation Ireland (SFI), Department of Business, Enterprise and Innovation (DBEI), Ireland; NWO, The Netherlands; The Science and Technology Facilities Council, UK; Ministry of Science and Higher Education, Poland. This research made use of the Dutch national e-infrastructure with support of the SURF Cooperative (e-infra 180169) and the LOFAR e-infra group. The J{\"u}lich LOFAR Long Term Archive and the German LOFAR network are both coordinated and operated by the J{\"u}lich Supercomputing Centre (JSC), and computing resources on the Supercomputer JUWELS at JSC were provided by the Gauss Centre for Supercomputing e.V. (grant CHTB00) through the John von Neumann Institute for Computing (NIC).

\item[Author contributions] L.G. coordinated the research, led the {\it XMM-Newton}, LOFAR
and GMRT proposals, reduced and analyzed the {\it Chandra} and {\it XMM-Newton} data, and wrote the manuscript.
H.A. led the {\it Chandra} proposal and analyzed the {\it Suzaku} data. T.W.S, H.T.I., R.J.vW.,
and F.G. performed the radio data reduction and worked on the radio spectral index map. F.M., J.M., I.U.,
and J.dP. assisted the interpretation of the X-ray results, V.P. assisted the GMRT proposal, 
H.J.A.R. and J.S.K. provided extensive suggestions on the manuscript.

\item[Competing interests] The authors declare that they have no
competing financial interests.
\end{addendum}

\newpage

\begin{figure*}[!htbp]
\begin{center}
\includegraphics[angle=0,width=0.75\columnwidth]{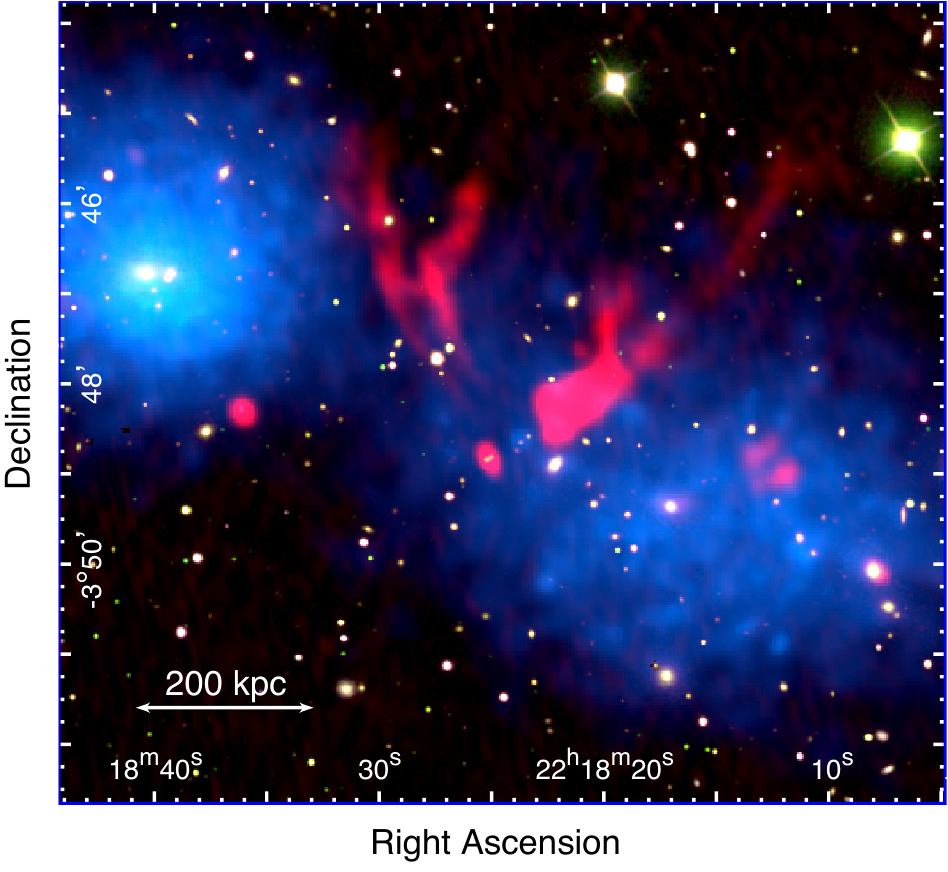}
\caption{\textbf{Composite image of the pre-merging cluster 1E~2216.0-0401 and 1E~2215.7-0404.} 
The SDSS $gri$ image is overlaid with the radio emission at 325~MHz from GMRT (red), and
$0.5-8.0$~keV X-ray emission from {\it Chandra} (blue). The linear scale is provided in the 
bottom left corner. }
\label{fig:composite}
\end{center}
\end{figure*}

\newpage 

\begin{figure*}[!htbp]
\begin{center}
\includegraphics[angle=0,width=\columnwidth]{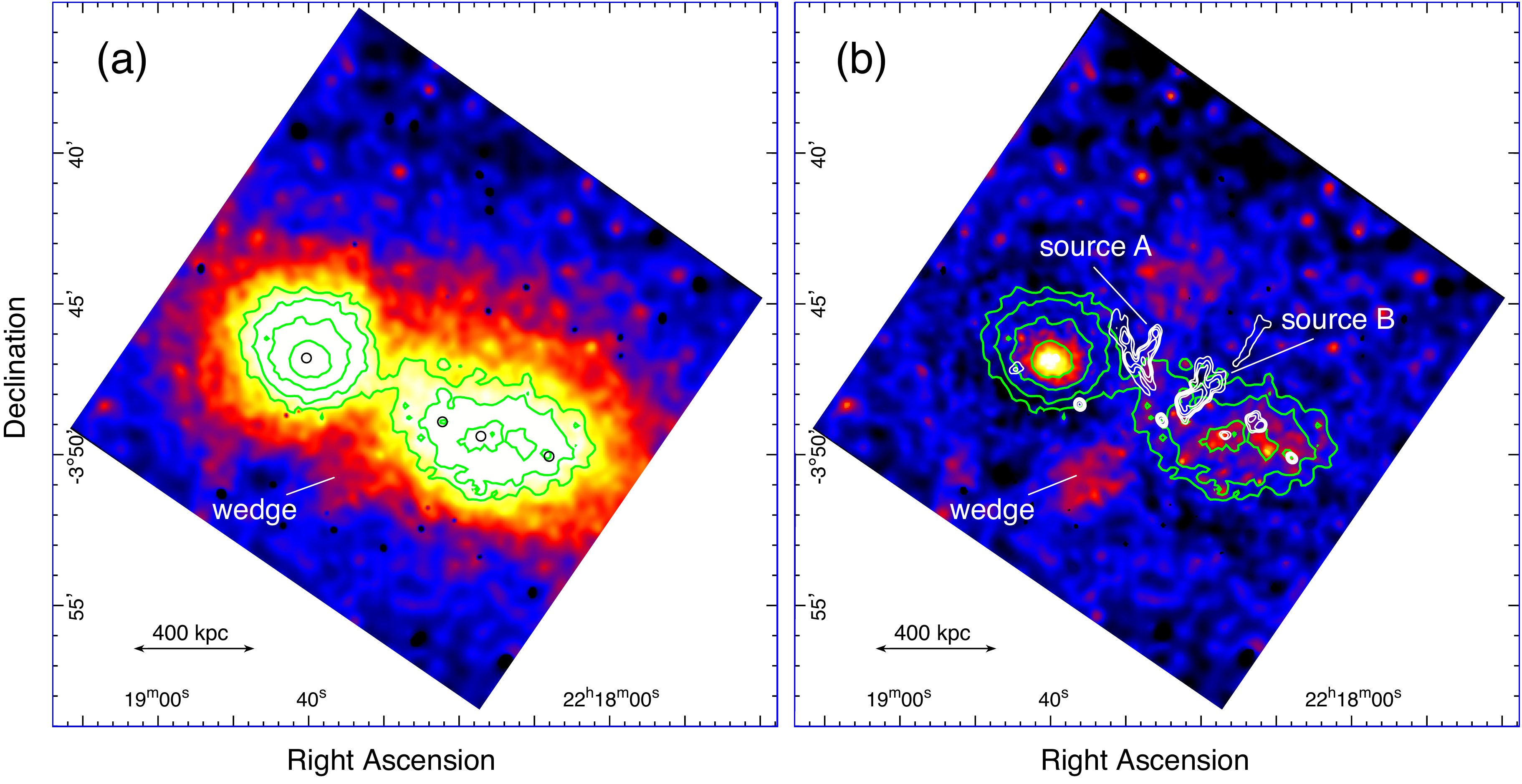}
\caption{\textbf{Original X-ray image and residual image from Chandra.} 
(a) {\it Chandra} image highlighting the protruding features towards the southeast on the
merger equator. Green contours show the X-ray distribution of the central regions. The contours
are drawn at levels of 2.5, 3.5, 4.5, and 8 times the local background. Positions of the 
dominant galaxies are marked with black circles. (b) Residual {\it Chandra} image, created by subtracting the
2-D beta-like model as shown in Supplementary Figure~1. The wedge-like feature is clearly seen in the residual image. 
The green and white contours show the X-ray (same as in panel a) 
and radio 325~MHz emission, respectively. The radio contours are drawn at levels of $\sqrt{[1,3,9,...]}$ $\times$ 4$\sigma_{\rm rms}$, where
$\sigma_{\rm rms}$ is the map noise. }
\label{fig:chimg}
\end{center}
\end{figure*}

\newpage 

\begin{figure*}[!htbp]
\begin{center}
\includegraphics[angle=0,width=\columnwidth]{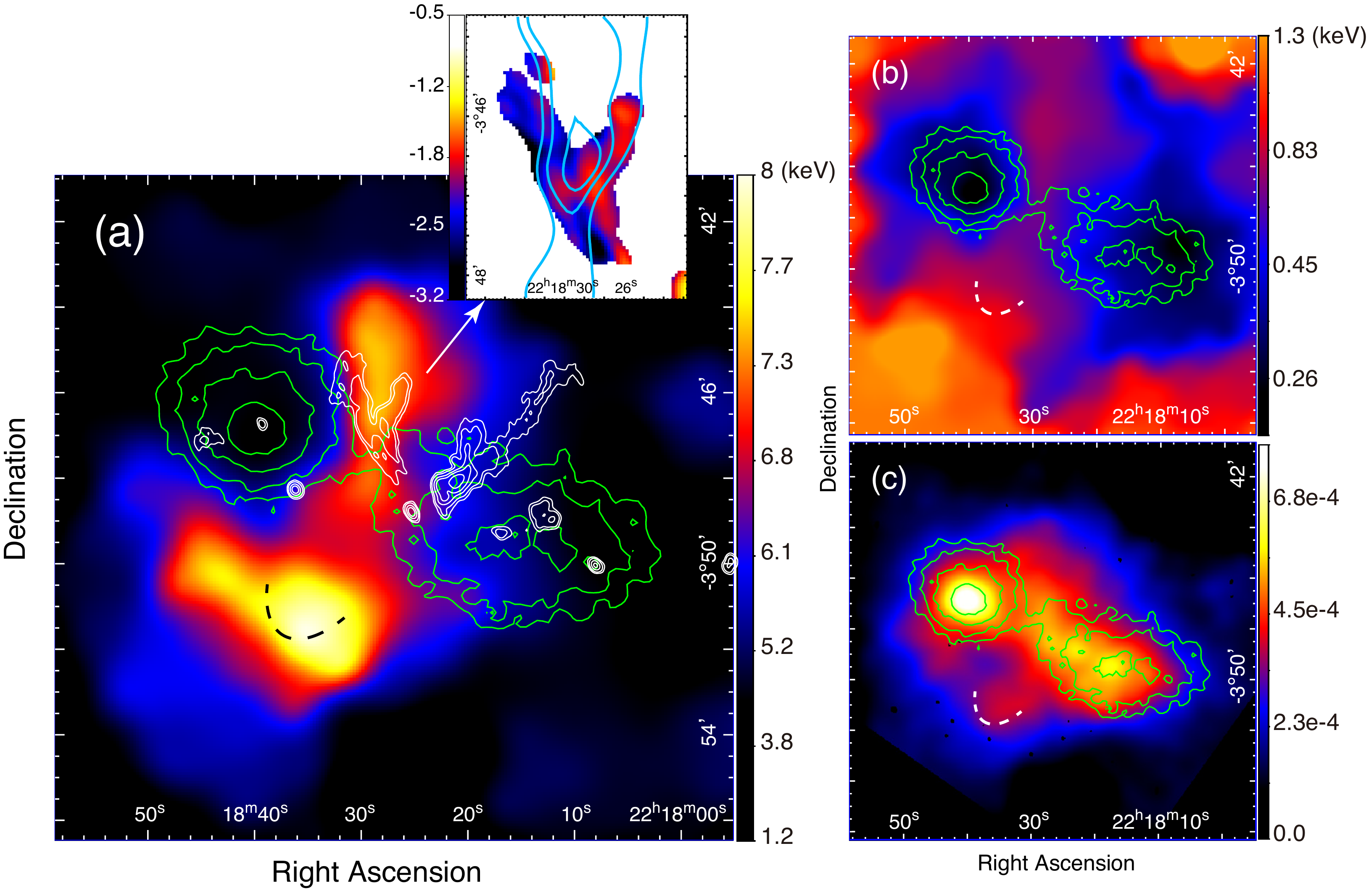}
\caption{\textbf{ICM temperature map, temperature uncertainty map, and pressure map of 1E~2216.0-0401 and 1E~2215.7-0404.} 
(a) {\it XMM-Newton} temperature map overlapped with contours from the X-ray (green) 
and radio 325~MHz (white)
images. Both the X-ray and radio contours are the same as in Figure~2. The edge of the wedge feature is marked with a black dashed curve. High ICM temperature is observed at the wedge and also on the opposite side of the merger axis. An inset shows the radio spectral index map derived from 144~MHz to 325~MHz of the radio tails (source A), overlapped with contours from the high ICM temperature structure (light blue). The contours in the inset are taken at levels of 6.8~keV, 7.1~keV and 7.4~keV. 
(b) Statistical error (1$\sigma$) of the temperature map. (c) Pseudo pressure map in arbitrary units, created from the {\it XMM-Newton} temperature map and a smoothed {\it Chandra} image. The white dashed curve shows the position of the wedge.}
\label{fig:xmmtmap}
\end{center}
\end{figure*}


\newpage

\begin{methods}

We assume that $\rm H_{0} = 70$ km s$^{-1}$ Mpc$^{-1}$, $\Omega_{\rm M} = 0.27$ and $\Omega_{\rm \Lambda} = 0.73$.
At the distance of the 1E2216 cluster, 1$^{\prime}$ corresponds to 91.8 kpc. All spectra are fitted using the C-statistics, and the quoted uncertainties are 1$\sigma$.

\begin{table}[h!]
\caption{\textbf{List of {\it Chandra}, {\it XMM-Newton}, and {\it Suzaku} observations used in this paper.} \label{tab:list}}
\begin{center}
\begin{tabular}{lcccc}
\hline
\hline
Instrument & Observation & Date & Initial exposure (ks) & Clean exposure (ks) \\
\hline
{\it Chandra} ACIS & 20778 & 2018-07-23 & 38.0 & 37.6 \\
{\it Chandra} ACIS & 21131 & 2018-07-25 & 32.1 & 31.7 \\
{\it Chandra} ACIS & 21132 & 2018-07-26 & 36.1 & 35.6 \\
{\it Chandra} ACIS & 21133 & 2018-07-27 & 25.6 & 25.3 \\
{\it Chandra} ACIS & 21134 & 2018-07-29 & 17.1 & 16.9 \\
\hline
{\it XMM-Newton} EPIC & 0800380101 & 2017-11-11 & 139.0 & 79.5 (MOS), 53.7 (pn) \\
\hline
{\it Suzaku} XIS & 807085010 & 2012-05-16 & 28.6 & 24.9 \\
{\it Suzaku} XIS & 807084010 & 2012-11-29 & 15.8 & 11.2 \\
\hline
\end{tabular}
\end{center}
\end{table}

 \subsection*{{\it Chandra} data processing}
 
{\it Chandra} observed 1E2216 on 2018 July $23-29$ for a total exposure of 149~ks with the ACIS-I detector, pointing at 
the bridge region between the two merging clusters. The datasets are reprocessed using the software {\it chandra\_repro} in
the latest CIAO 4.10 package. Light curves are extracted using 200~s time bins in the $0.5-12.0$ keV for the point-source-free
region. These light curves show no flares, therefore the entire observing time period is used. For each observation the bright point 
sources are removed using the tool {\it celldetect} with the detecting threshold of 3$\sigma$ on the $0.5-8.0$ keV image. The image 
is further checked by eye for fainter point sources.

The X-ray background is removed using the blank sky background dataset. For each observation, the tool {\it blanksky} is run 
to locate the blank sky files, and to reproject the blank sky to match the observation. The background files are reprocessed to 
have the same gain, timing, and bad pixel properties as the observation data.

The exposure map is created for each observation using the tool {\it mkexpmap} in the $0.5-8.0$ keV band. To take into 
account the energy dependence, we provide a model spectrum to {\it mkexpmap}, which has a temperature of 5 keV, abundance of 
0.3 solar, and a Galactic absorption column of $7.3 \times 10^{20}$ cm$^{-2}$ \cite{willingale2013}. 
The edges of the detector are excluded from the 
following analysis, as they show some possible artefacts caused by the exposure correction. The final flux image is then created 
by removing the background from the original image, and dividing it by the exposure map. Figure~2 shows the flux image 
in the $0.5-8.0$ keV. It is smoothed by a Gaussian kernel with an adaptive scale to include about 50 net counts in each kernel.

To highlight the possible surface brightness excess on the equatorial plane of the merger, we subtract the X-ray emission from the main haloes of the two clusters. The 
classical beta profile 
is used to model the projected 2-D gas distribution in the $0.5-8.0$ keV band 
for each cluster. The equatorial
regions, as shown in Supplementary Figure~2(a), are excluded in the fits. The shape of the King profile, the central position and 
the ellipticity of the X-ray haloes, are all set free in the fits. The best-fit model is then 
smoothed by a Gaussian kernel with the same scale as is done to the flux image. As shown in 
Supplementary Figure~1, the resulting beta-like image is used to calculate the residual map (Figure~2).

For the spectral analysis, the source spectral files are extracted separately from each of the observations, 
and the responses and ancillary response matrices are created using {\it mkacisrmf} and {\it mkwarf}, respectively. 
Background spectra are extracted from the blank sky datasets. The spectra of the same region are 
fit simultaneously between 0.5~keV and 7.0~keV using the {\it SPEX} package with SPEXACT 3.04.
The raw spectra are re-grouped with the optimal binning
method provided by the {\it SPEX} package.

To obtain an overview of the two clusters, we analyse two spectra, each covering the central 350~kpc region of one cluster.
The {\it Chandra} spectra are fitted with a single redshifted {\it cie} model for the emission measure, temperature, average 
metallicity, and the redshift.
The proto-Solar abundances from \cite{lodders2009}
are adopted. A foreground Galactic absorption column of $N_{\rm H} = 7.3 \times 10^{20}$ cm$^{-2}$ is
included in the fits. The best-fit temperatures are $4.6 \pm 0.1$~keV and $5.4 \pm 0.1$~keV, the average metallicities are 
$0.36 \pm 0.06$ solar and $0.28 \pm 0.05$ solar, and the redshifts are $0.095 \pm 0.004$ and $0.091 \pm 0.004$, 
for the northeast and southwest clusters, respectively. The obtained parameters agree well with those reported in the previous 
{\it Suzaku} work. Based on the mass-temperature relation found in \cite{vik2009}, the approximate total masses
of the two clusters are $2.5 \times 10^{14}$ solar mass and $3.2 \times 10^{14}$ solar mass.

\subsection*{{\it XMM-Newton} data processing}

{\it XMM-Newton} observed 1E2216 on 2017 November 11-13 for a total exposure of 139~ks. The latest XMM science analysis
system (SAS) and the extended source analysis software (ESAS) are used to process and calibrate the data obtained with
the European Photon Imaging Camera (EPIC) onboard XMM. The tools {\it emchain} and {\it epchain} are used to create the 
reduced EPIC-MOS and pn event files, respectively. Light curves are extracted in 100~s bins and screened for background flares
with the {\it mos-filter} and {\it pn-filter} tasks. The final exposures reduce to $\sim 80$~ks for the MOS and $\sim 54$~ks for the pn data.
The events with PATTERN values greater than 12 (MOS) / 4 (pn) and non-zero FLAG values are also excluded. The exposure-corrected 
images are made in the 0.5 to 2 keV band using a bin size of 40 pixels.
The MOS image is used to identify point sources from the diffuse components. The sources detected above a flux threshold 
of $5 \times 10^{-14}$ ergs cm$^{-2}$ s$^{-1}$ are excluded in the analysis. The {\it Chandra} image is used to cross-check the
point source detection made with the MOS.

{\it XMM-Newton} provides a better statistics of the target cluster than {\it Chandra} and {\it Suzaku}. 
The source spectra, particle background, and response files are prepared by the {\it mos-spectra} and 
{\it pn-spectra} tools.
The remaining sky background, including the foreground Galactic soft X-ray emission and the unresolved cosmic
X-ray background (CXB) are taken into account as spectral components in the fits. The Galactic emission is modelled by
two thermal plasma components (the SPEX {\it cie} model) with fixed temperatures of 0.08~keV and 0.3~keV. The 0.08~keV component 
is free of absorption while the 0.3~keV one is absorbed by the Galactic ISM. The abundances and redshifts of the two components
 are fixed to unity and zero, respectively, while their normalizations are set free to vary. The CXB component is represented
 as an absorbed power law, with a fixed photon index of 1.4 and flux of $6.4 \times 10^{-8}$ ergs cm$^{-2}$ s$^{-1}$ sr$^{-1}$ \cite{gu2012}. 
 
We examine the {\it XMM-Newton} spectra from the ICM properties in the central 350~kpc regions of the two clusters.
The spectra are re-grouped with the optimal binning method. 
The ICM component is modelled by one redshifted {\it cie} component,
absorbed by the Galactic neutral materials. The column density is fixed to $7.3 \times 10^{20}$ cm$^{-2}$. 
The temperature, average metallicity, redshift, and emission measure of the thermal component are left free in the fits. 
The best-fit temperatures are $4.7 \pm 0.1$~keV and $5.1 \pm 0.1$~keV, and the average metallicities are
$0.42 \pm 0.04$~solar and $0.33 \pm 0.03$~solar, for the northeast and southwest clusters, respectively. 
The redshifts of the two objects are $0.094 \pm 0.001$ and $0.090 \pm 0.001$. These values are consistent with those
determined with the {\it Chandra} spectra.

\subsection*{{\it Suzaku} data processing}

The \textit{Suzaku} X-ray Imaging Spectrometer (XIS) observed 1E2216.0-0401 and 1E2215.7-0404 on May and November 2012 
for exposures of 28.6 and 15.8~ks, respectively. 
Data reduction is performed with HEAsoft version 6.24 and the latest CALDB version (20160607). 
To suppress instrumental background, an event screening with the cosmic-ray cutoff rigidity COR2 $>$ 6 GV is applied. 
Further reductions are applied to the XIS1 to mitigate background increase due to the change of the charge injection (\url{http://www.astro.isas.jaxa.jp/suzaku/analysis/xis/xis1\_ci\_6\_nxb/}).
Because of the damage from a micro-meteoroid strike, XIS2 is not in use. The resultant clean exposures are about 
25 and 11 ks for 1E2216.0-0401 and 1E2215.7-0404, respectively.

For the spectral analysis, we model the sky background in the same way as is done for {\it XMM-Newton}. All the XIS spectra
are fit simultaneously. 


\subsection*{Radio data preparation}

Radio observations at 235, 325, 610 and 1300~MHz were obtained with the GMRT under project codes 28\_087 and 31\_079. 
At 235 and 610~MHz (dual-frequency observing), the target was observed on 2015-08-19 for 2.2~hours over an effective 
bandwidth of 18 and 33~MHz, respectively. At 325 MHz, the target was observed on 2016-12-24 for 6.1~hours over an 
effective bandwidth of 33~MHz. At 1300 MHz,  the target was observed on 2016-12-23 for 2.8~hours over an effective 
bandwidth of 33~MHz. All data was processed using the SPAM pipeline,
which included flux scale and bandpass calibrations using 3C\,48, and multiple rounds of direction-dependent calibration and imaging. The final images have sensitivities of 570, 130, 57, and 96 $\mu$Jy/beam at 235, 325, 610, and 1300~MHz, respectively. The resolutions are 14.8'' x 9.6'', 11.7'' x 8.0'', 6.5'' x 3.8'', and 2.4'' x 1.9'', respectively. The 1300~MHz image is not used in this paper, because it is 
not deep enough to detect the main part of the target source.

The LOFAR high band antenna (HBA) data were obtained as part of projects LC7\_003 and LC9\_014 and the target was observed for 4~hours on four separate occasions between 2017-01-19  and 2017-11-21 with a frequency coverage of 120-168~MHz. All the data were passed through the standard direction independent calibration pipeline, 
and made use of the LOFAR Long Term Archive (LTA) compute facilities for efficient processing. 
The data were then processed through the latest version of the LOFAR Two-metre Sky Survey (LoTSS) 
direction dependent calibration pipeline \cite{tim2019}. This pipeline uses {{k\sc{ms}}} 
\cite{smirnov2015} 
to calibrate for ionospheric effects and errors in the beam model and {{{\sc ddf}acet}} 
\cite{tasse2018} 
to apply the derived solutions during the imaging. To further enhance the quality of the images a post processing step was added to remove all sources from the field besides the target and to self calibrate the data to improve the accuracy of the calibration solutions at that location. The final data has a sensitivity of 0.27~mJy/beam in the vicinity of the target when imaged with a resolution of 17.6'' x 6.3''.






\subsection*{Thermal structure of the pre-merger system}

To obtain a global view of the thermal structure, we map the two-dimensional distribution of the ICM temperature in the early merger cluster. The map shown in Figure~3 is made based on the method described in \cite{gu2009}. First, we randomly select a set of pixels to be the centers of extraction regions. The distribution of ``knots'' is sufficiently dense, with a separation $< 10^{\prime\prime}$ between any two adjacent knots, to ensure that the map is fully covered by the extraction regions. 
For each knot, a spectrum is extracted from a circular region
containing a net counts of about 7000 (or a signal to noise of 84) in the $0.5-7.0$ keV band. The background is corrected
with the {\it XMM-Newton} blanksky. Typical radii of the
extraction regions are $\sim 0.3^{\prime}-1.5^{\prime}$. The best-fit temperatures, determined from the fits with the SPEX {\it cie} model, are assigned to the corresponding knots. The temperatures of the remaining pixels are obtained through interpolation. 
For a pixel at position $(x,y)$, we define a circular region containing 7000 counts, with a radius of $r(x,y)$. The 
temperature at $(x,y)$ is then calculated through a weighted average over all the knots $(x_{\rm k}, y_{\rm k})$ in the circular region, 
\begin{equation}
T(x,y) = \sum_{\rm k} G_{\rm k}(R) T (x_{\rm k}, y_{\rm k}) / \sum_{\rm k} G_{\rm k}(R),  
\label{eq:tmap}
\end{equation}
where $R$ is the distance from $(x,y)$ to $(x_{\rm k}, y_{\rm k})$, and $G_{\rm k}$ is the Gaussian kernel with the scale parameter $\sigma$ fixed to $r(x_{\rm k},y_{\rm k})$. The use of Gaussian function produces a smooth temperature map, 
whose resolution is set by the extraction region. As the main body of the map is essentially derived through interpolation, it is little
affected by the bias from the possible poor fits in individual regions.

The statistical errors on the temperature are calculated with the {\it error} command in the fits. In the fits of a few regions,
the error calculation could lead to a smaller C-statistic than the original fits, we update the fits with the new
value and run the error calculation again. The temperature errors map is derived through the same technique 
as the temperature map. As shown in Figure~3, the temperature errors are typically 
$\sim 0.3$ keV for the cluster central regions, $\sim 0.6$ keV for the bridge region between the two clusters,
and $\sim 0.8$ keV at the southeast shock region. The errors are about 10\% of the values in the temperature map, except
for the outermost regions at $r \geq 7^{\prime}$, where the errors exceed 20\% of the local temperatures.

As a crosscheck, we further calculate the temperature maps using the standard weighted Voronoi Tesselations (WVT) algorithm \cite{diehl2006, simi2007, dp2010, og2013, su2017} (\url{http://www.phy.ohio.edu/~diehl/WVT/})
and the adaptive circular binning (ACB) \cite{randall2008, clarke2013}. 
The WVT binning divides the map into 150 non-overlapping regions, each 
contains 7000 clean counts in $0.5-7.0$ keV. The temperatures between two adjacent WVT bins are determined independently. The ACB method extracts spectra from a circular region, which contains 7000 clean counts, centered on each ``pixel'' as shown in Supplementary Figure~6(c). The ACB regions significantly overlap with each other in the cluster outskirt. As can be seen in Supplementary Figure~6, the three temperature maps obtained through different methods agree well with each other. For most of the regions, the discrepancies among the maps are found to be $\leq 0.5$ keV, smaller than the typical statistical errors from the 
spectral fits. 

As shown in Figure~3, the peak of ICM temperature can be seen at the southeast wedge, probably due to the 
heating by the equatorial shock. The temperature
decreases, from $\sim 8$~keV to $\sim 6$~keV, across the surface brightness discontinuity. It seems that the temperature gradient
in the map is slightly milder than the deprojected temperature profile for the post-shock and pre-shock regions, which reveals a jump from
$8.1$~keV to $5.2$~keV. This is probably because that hot spot in the post-shock region
has been blurred by the interpolation method (Eq.~\ref{eq:tmap} in the Method section).

Combining the {\it Chandra} image and the {\it XMM-Newton} temperature map, we can further obtain the approximate 
distributions of gas pressure and entropy. The electron number density distribution can be directly determined 
from the X-ray surface brightness $S$, as 
\begin{equation}
S \approx \Lambda(T,A) \int n_{\rm e}^{2} dl,
\end{equation}
where $\Lambda(T,A)$ is the cooling function that depends on the ICM temperature and metallicity, and the electron 
density $n_{\rm e}$ is integrated along the line of sight. Ignoring the cooling function will introduce a 
uncertainty of about 10\% in the pressure and entropy maps, which is similar to the typical error of the temperature map.

By approximating electron density as $S^{1/2}$, the pseudo-pressure and pseudo-entropy are then defined as
\begin{equation}
P = T S^{1/2},
\end{equation}
\begin{equation}
K = T S^{-1/3}.
\end{equation}
The {\it Chandra} $0.5-2.0$ keV image is used for the pressure and entropy maps. The {\it Chandra} image
is smoothed with the {\it XMM-Newton} point spread function, and then convolved with the 
the same adaptive smoothing kernel $r(x,y)$ as the {\it XMM-Newton} temperature map. A region of both 
high pressure and entropy might indicate the presence of a shock. As shown in Figure~3 and Supplementary Figure~7, the southeast
shock is indeed shown as a high pressure and entropy region, while such a dramatic structure is not visible
at the northwest. The high pressure structure nicely traces the southeast shock, and shows an abrupt drop outside
the front. Reading directly from the map, the ratio of the post-shock pressure $P_{\rm post}$ to pre-shock pressure 
$P_{\rm pre}$ is $\sim 2.4$, and the shock Mach number can be estimated as
\begin{equation}
\mathcal{M} = \sqrt[]{\frac{(\gamma + 1)\frac{P_{\rm post}}{P_{\rm pre}}+(\gamma-1)}{2\gamma}} \approx 1.5.
\end{equation}
It agrees with the values obtained from the surface brightness and temperature jump analysis. Such a gas pressure jump
can potentially be confirmed with a dedicated high spatial resolution observation of the Sunyaev-Zel'dovich effect \cite{korngut2011, yamada2012, erler2015, basu2016}. The pressure map also reveals a mild gradient at the north, extending up to $\sim 3^{\prime}$ from the collision axis.

As shown in Supplementary Figure~7, the high entropy feature at the southeast appears to extend to a larger radius than the detected 
shock front, even taking into account the relatively low resolution of the temperature map. This might
indicate the presence of other ICM heating
processes in the past, such as previous equatorial shocks and/or adiabatic compression by the merger. The high entropy is also seen along the equatorial direction 
at the north 
of the collision axis, although the average gas entropy in the north is lower than the value in the south.

\subsection*{Data availability}
The data that support the plots within this paper and other findings of this study 
are available from the corresponding author upon reasonable request.

\end{methods}


\setcounter{figure}{0} 
\section*{Supplementary Information}

\subsection*{Images for shock detection in X-ray surface brightness}

\renewcommand{\figurename}{Supplementary Figure}
\begin{figure*}[!htbp]
\begin{center}
\includegraphics[angle=0,width=0.5\columnwidth]{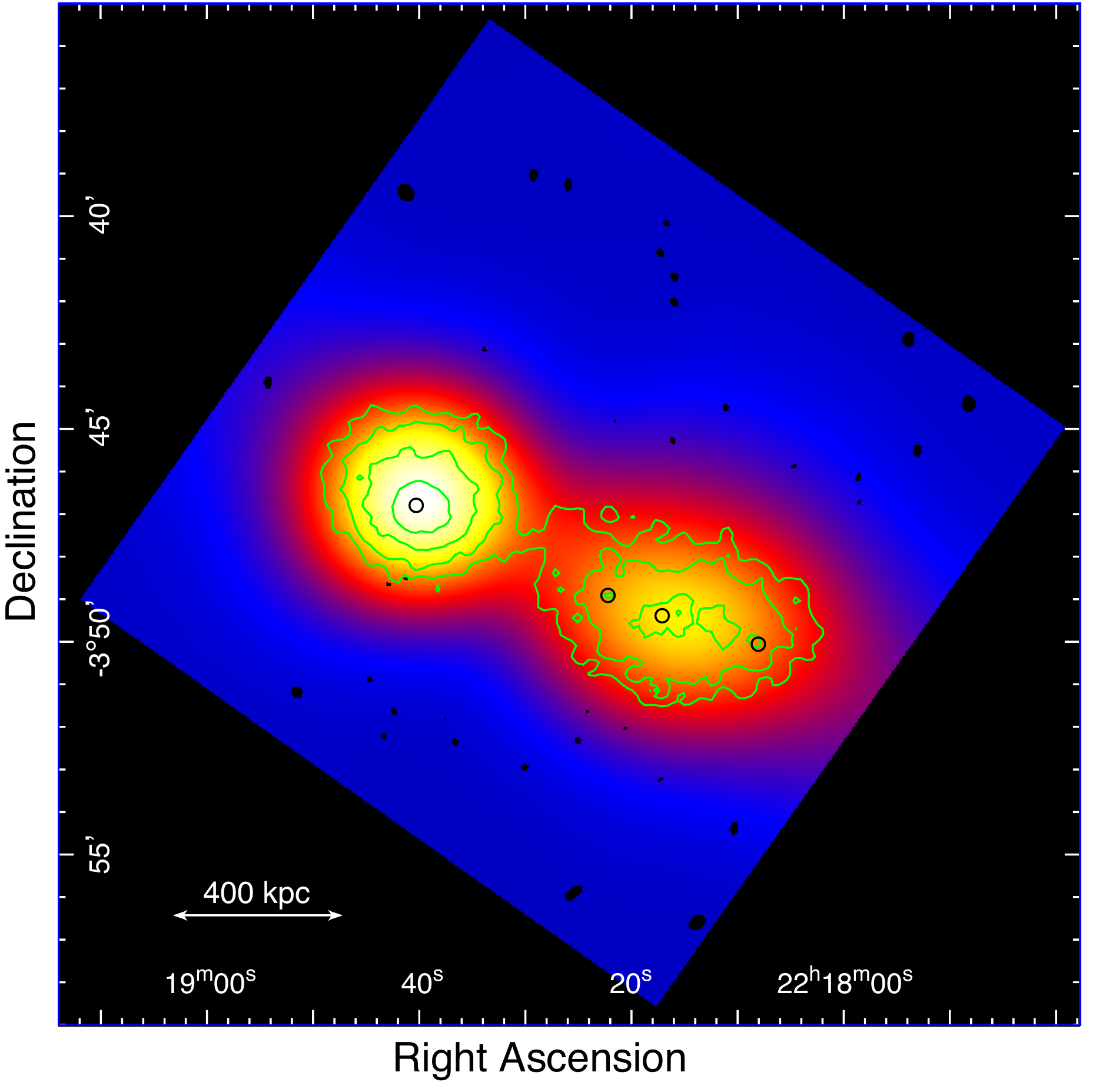}
\caption{\textbf{Beta model fits to the {\it Chandra} image.}
This model image is produced by fitting the surface brightness in the $0.5-8.0$~keV with a 2-D beta model.
It is used to create the residual map in Figure~2. Green contours show the X-ray brightness of the 
central region. The contour levels are the same as in Figure~2. Positions of the dominant galaxies are marked with black circles. }
\label{fig:beta}
\end{center}
\end{figure*}

\begin{figure*}[!htbp]
\begin{center}
\includegraphics[angle=0,width=0.8\columnwidth]{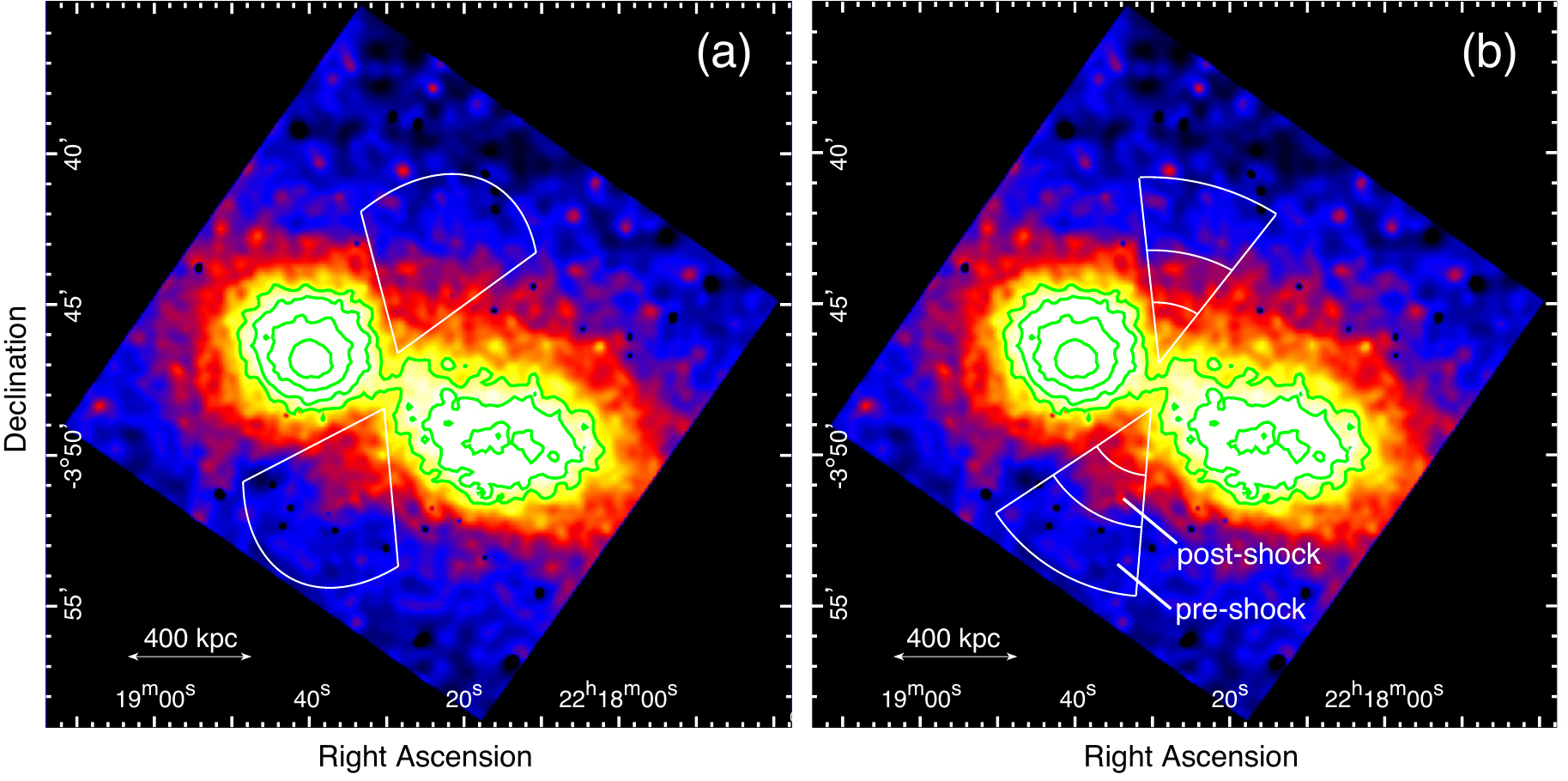}
\caption{\textbf{Regions used in the shock detection.}
Regions with white solid lines on the {\it Chandra} images are used 
for extracting (a) the surface brightness profiles (Supplementary Figures~\ref{fig:sbprof} and \ref{fig:xmmsbprof}) and (b) the 
X-ray temperature profiles (Supplementary Figure~\ref{fig:shockkt}). Green contours show the X-ray brightness of the 
central region. The contour levels are the same as in Figure~2. }
\label{fig:chreg}
\end{center}
\end{figure*}

\begin{figure*}[!htbp]
\begin{center}
\includegraphics[angle=0,width=0.75\columnwidth]{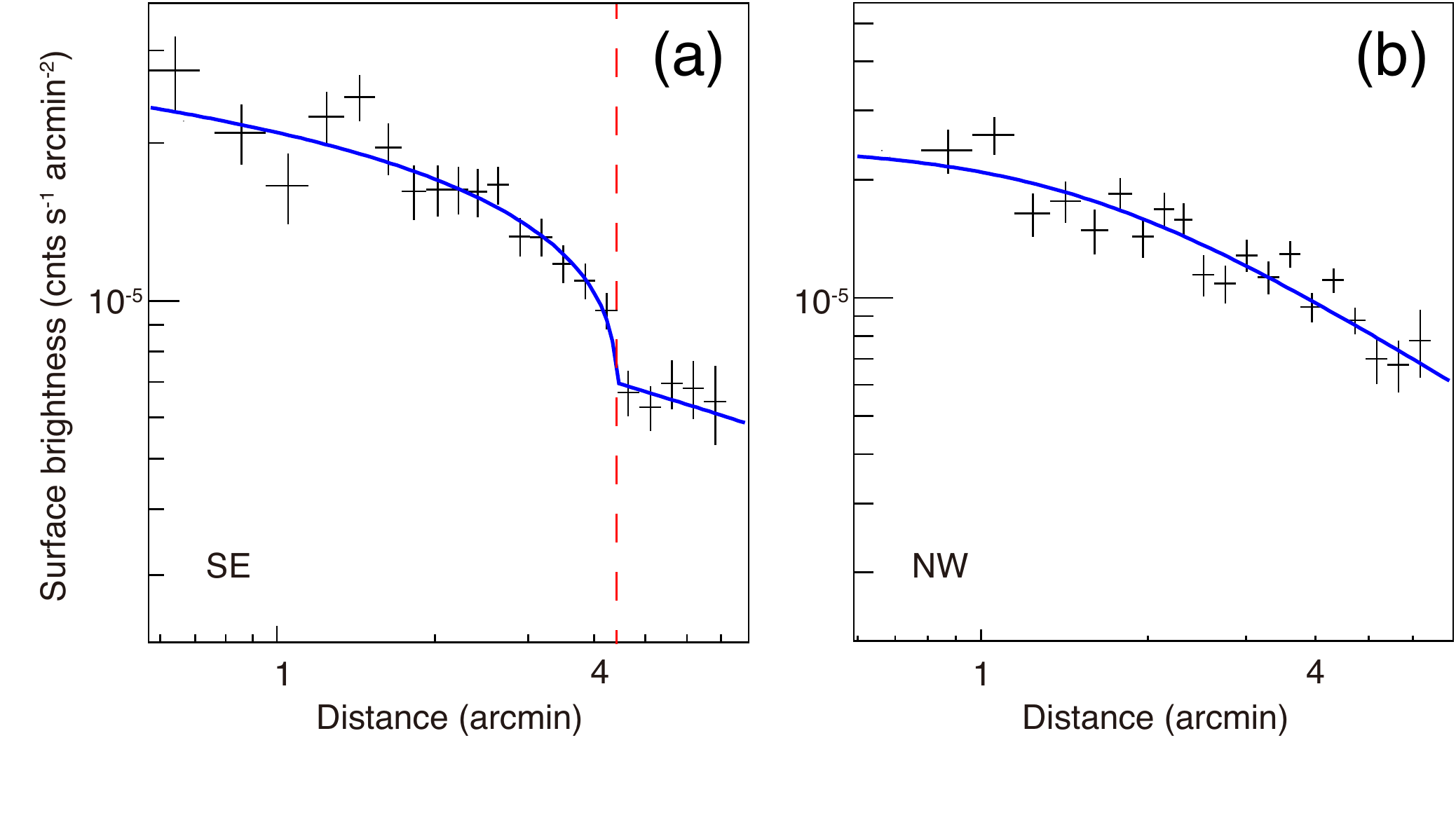}
\caption{\textbf{Chandra surface brightness profiles along the merger equator.}
(a) {\it Chandra} surface brightness profile in the $0.5-8.0$ keV band across the SE equatorial shock in an elliptical sector. 
(b) Surface brightness profile in the NW region. Errors are given at the 1$\sigma$ level.
The blue line in panel a shows the best-fit broken power law model with the compression
factor $C=2.1$, and the blue line in panel b shows the best-fit $beta$ model. The vertical red dashed line shows the surface brightness jump at a distance of $\sim $4.3$^{\prime}$ from the collision axis. }
\label{fig:sbprof}
\end{center}
\end{figure*}

\begin{figure*}[!htbp]
\begin{center}
\includegraphics[angle=0,width=0.75\columnwidth]{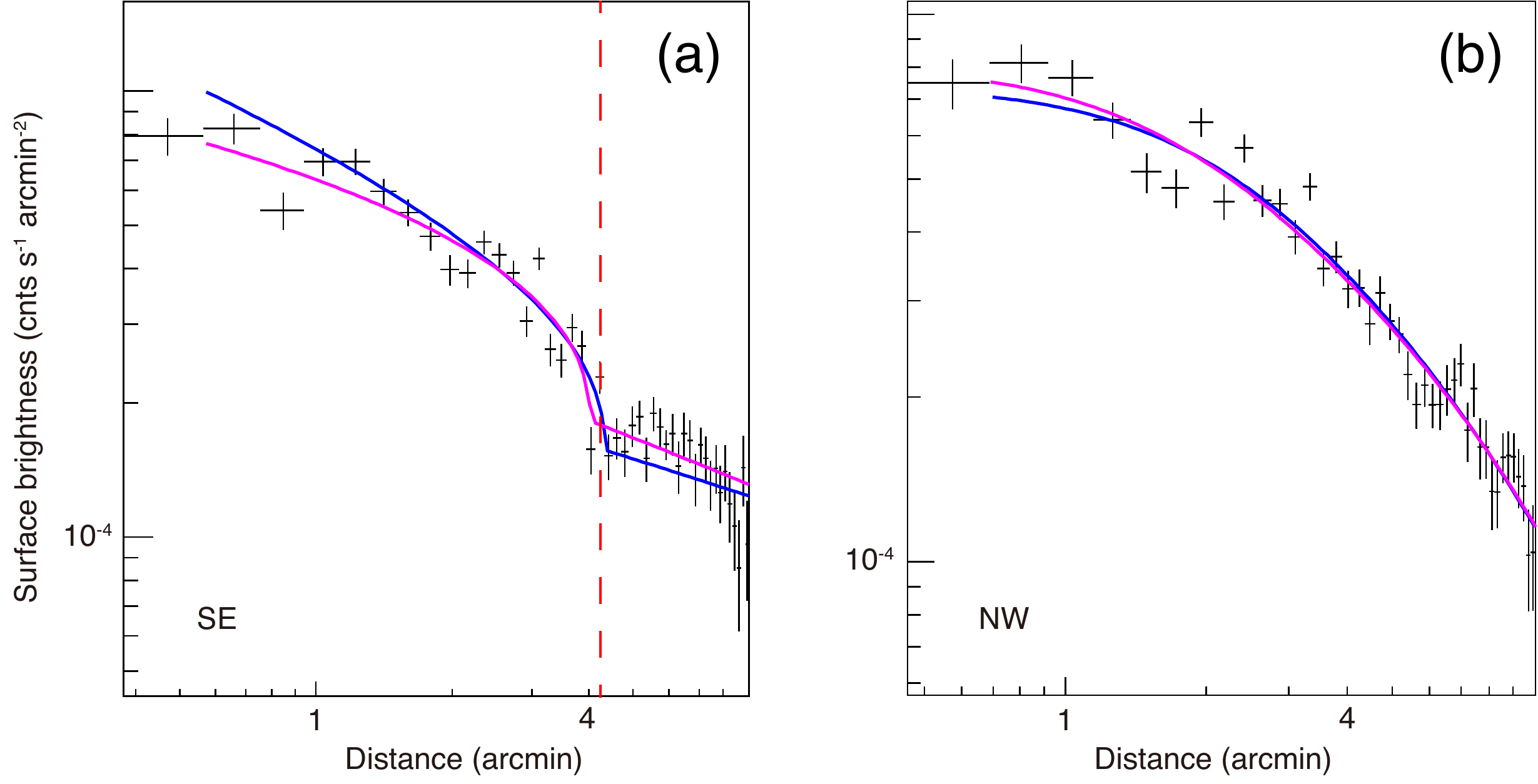}
\caption{\textbf{XMM-Newton surface brightness profiles along the merger equator.}
(a) {\it XMM-Newton} surface brightness profile in the $0.5-8.0$ keV band across the SE equatorial shock in an elliptical sector. (b) The surface brightness profile in the NW region. Errors are given at the 1$\sigma$ level. Magenta lines show the best-fit models to the {\it XMM-Newton} profiles,
and the blue lines show the {\it Chandra} models from Supplementary Figure~\ref{fig:sbprof}. The vertical red dashed line shows the distance of $\sim $4.1$^{\prime}$ from the collision axis.}
\label{fig:xmmsbprof}
\end{center}
\end{figure*}
\FloatBarrier

\subsection*{Images for shock detection in X-ray temperature}

\begin{figure*}[!htbp]
\begin{center}
\includegraphics[angle=0,width=0.8\columnwidth]{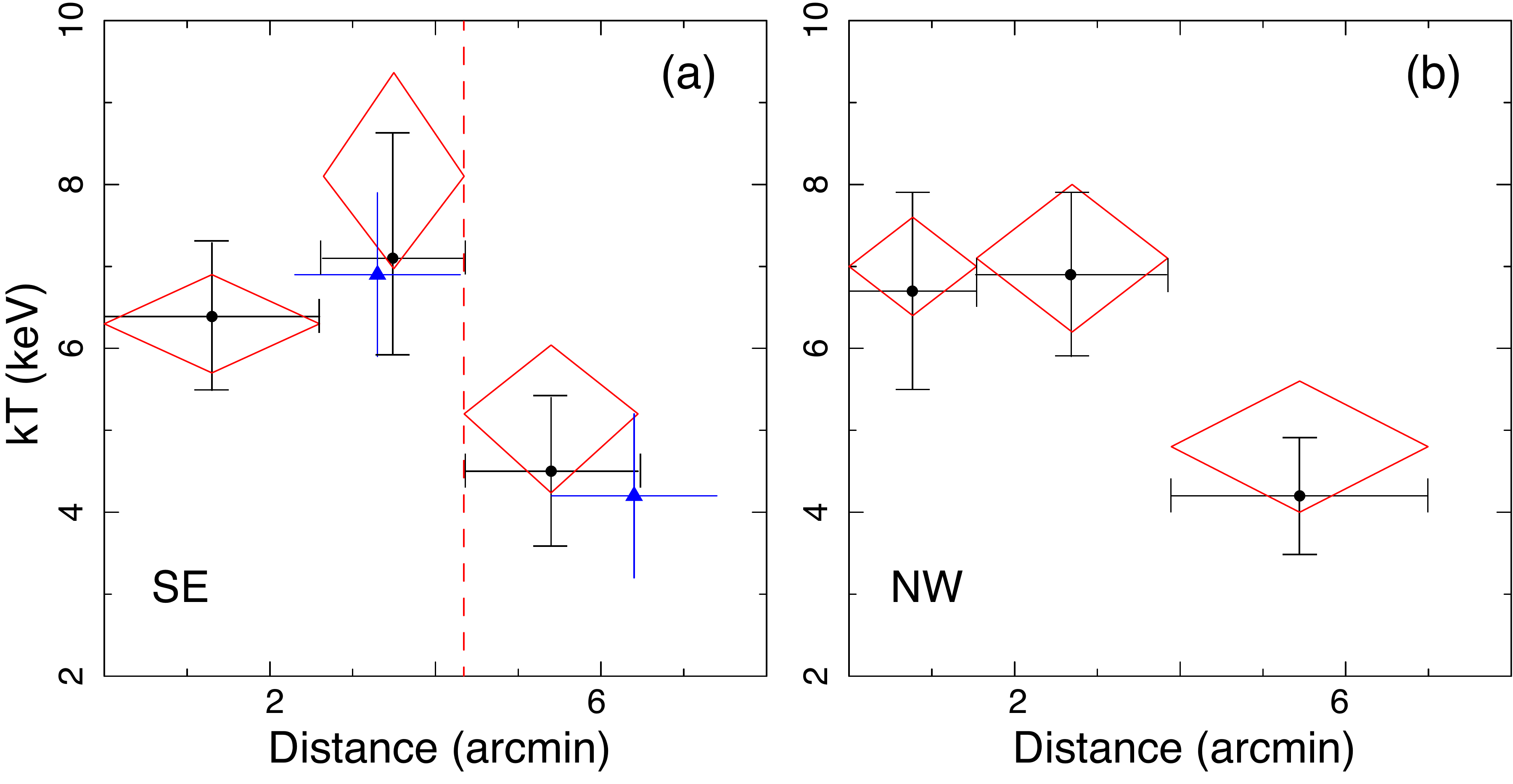}
\caption{\textbf{X-ray temperature profiles along the merger equator.}
(a) Projected temperature profiles obtained with {\it Chandra} (black) and {\it Suzaku} (blue), and 
deprojected profile with {\it XMM-Newton} (red), for the SE equatorial region. The vertical line 
marks the location of the surface brightness jump.
(b) Projected temperature profile from {\it Chandra} (black) and deprojected profile from
{\it XMM-Newton} (red) for the NW equatorial region. Errors are given at the 1$\sigma$ level.  }
\label{fig:shockkt}
\end{center}
\end{figure*}

\begin{figure*}[!htbp]
\begin{center}
\includegraphics[angle=0,width=0.95\columnwidth]{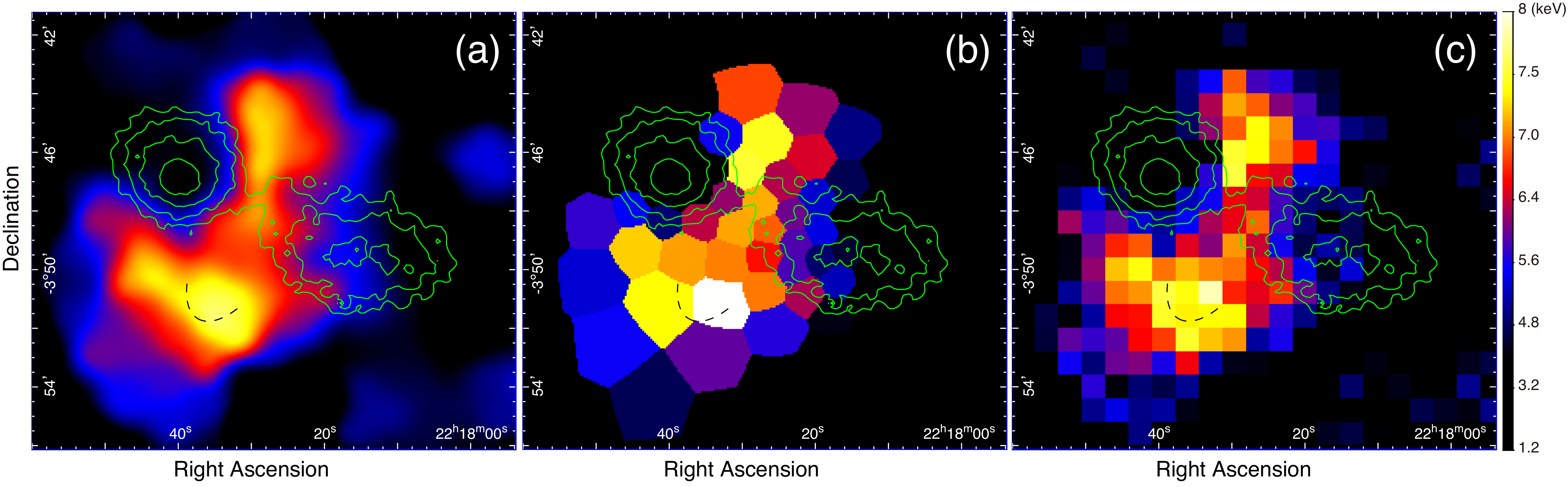}
\caption{\textbf{Temperature maps of 1E~2216.0-0401 and 1E~2215.7-0404 with three different algorithms.}
Temperature map obtained through (a) the fits and interpolation (Eq. 1 in the Method section) 
are compared with the maps 
obtained through (b) the WVT algorithm and (c) the ACB method. Green contours show the X-ray distribution (same as in Figure~2), and the black dashed curve indicates the position of the 
wedge.}
\label{fig:xmmtmap3bins}
\end{center}
\end{figure*}

\begin{figure*}[!htbp]
\begin{center}
\includegraphics[angle=0,width=0.6\columnwidth]{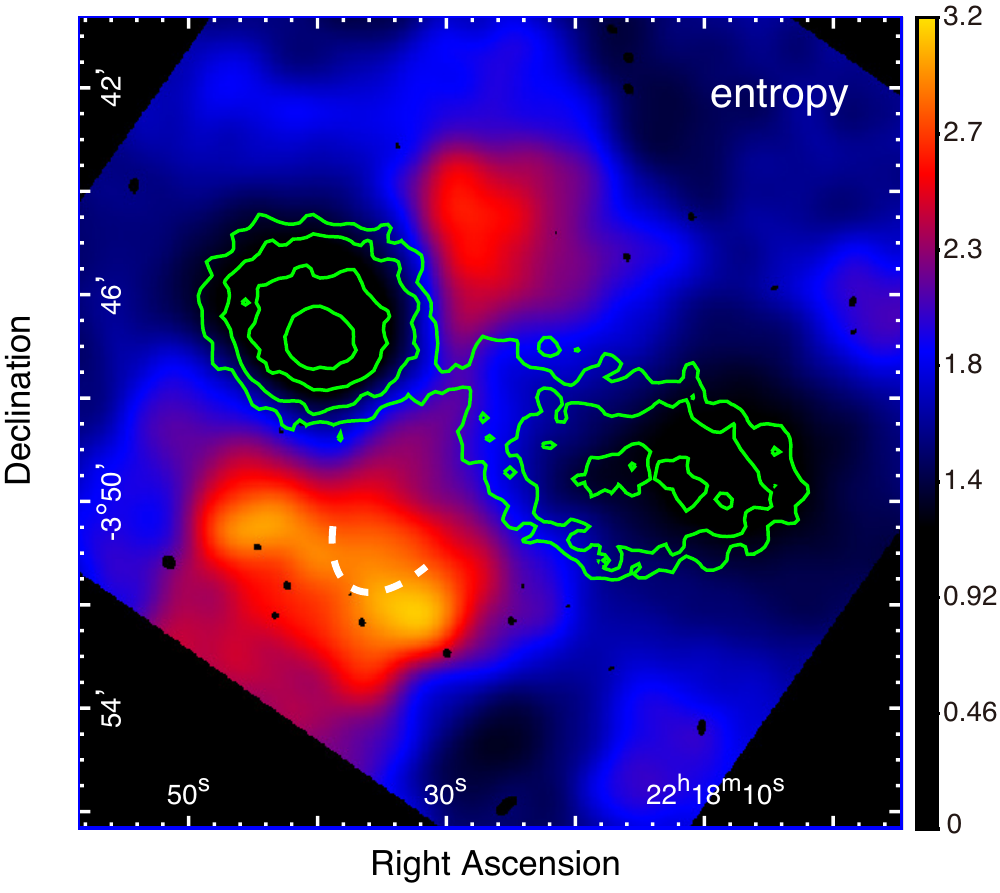}
\caption{ \textbf{Distribution of ICM entropy in the merger.}
Pseudo entropy map in arbitrary units, created from the {\it XMM-Newton} temperature map
(Figure~3) and a smoothed {\it Chandra} image. Green contours show the X-ray distribution (same as in Figure~2), and the white curve indicates the 
edge of the wedge-like feature.}
\label{fig:entropy}
\end{center}
\end{figure*}
\FloatBarrier

\subsection*{X-ray surface brightness profile of the equatorial shock}

To verify the wedge-like structure seen in the flux image, we study the {\it Chandra} surface brightness profile along the equatorial direction
of the merger. To avoid the contamination from the cluster main body, we extract the profiles in two elliptical sector regions (Supplementary Figure~2), one towards southeast and
another towards northwest, with an opening angle of 70 degree and a bin size of about $10^{\prime\prime}$. 
To fit the observed shape of the wedge, we set the major axes of the elliptical sectors on 
the equatorial plane. As shown in Supplementary Figure~3, the surface brightness profile is calculated 
as a function of the mean distance from the collision axis, which is defined as a straight line connecting the
X-ray peaks of the two clusters. The profiles 
are fit with the latest version of Proffit \cite{eckert2011} (\url{https://www.isdc.unige.ch/~deckert/newsite/Proffit.html}).

The surface brightness in the northwest sector decreases smoothly with the distance from the collision axis. It can be described by a standard $\beta$ model, with a best-fit $\beta = 0.31 \pm 0.03$
and core radius $r_{\rm c} = 1.5^\prime \pm 0.5^\prime$. The southeast region shows a clear jump at around 4.3$^{\prime}$, 
which appears to agree well with the shock front detected by eye. The profile is modelled with a 
broken power-law density distribution,

\begin{equation}
n(R) = \left\{
\begin{array}{c l}	
     C n_{0} R^{-\alpha_1} & R \leq R_{\rm edge}, \\
     n_{0} R^{-\alpha_2} & R > R_{\rm edge},
\end{array}\right.
\label{eq:bknpow}
\end{equation}
where $R$ is the 3-D radius, $C$ is the compression factor of the shock, $n_0$ is the normalization,
$\alpha_1$ and $\alpha_2$ are the power law indexes for the up and downstream regions, respectively,
and $R_{\rm edge}$ is the position of the jump. To project the 3-D density model into 2-D,
we assume that the shock surface is a doughnut-like structure around the merger axis, and the 
wedge-like surface morphology holds for the other parts of the doughnut. 


The projected broken power law model provides a reasonable fit to the southeast profile. The best-fit
power law indexes $\alpha_1$ and $\alpha_2$ are $0.35 \pm 0.05$ and $0.62 \pm 0.04$, respectively, the
cut-off radius $R_{\rm edge} = 4.3^\prime \pm 0.1^\prime$, and the compression factor $C = 2.1 \pm 0.5$.
The shock Mach number $\mathcal{M}$ can be obtained from the compression factor,
\begin{equation}
\mathcal{M} = \sqrt[]{\frac{2C}{\gamma + 1 - (\gamma - 1)C}} = \sqrt[]{\frac{3C}{4-C}},
\end{equation}
where the rightmost equation refers to the adiabatic index $\gamma = 5/3$ for the thermal plasma. 
By error propagation, the uncertainty in the Mach number can be calculated as 
\begin{equation}
\delta\mathcal{M} = \frac{6}{ (4 - C)^2 \mathcal{M}} \delta C
\end{equation}
Therefore, the 
Mach number estimated from the {\it Chandra} surface brightness profile is $1.8 \pm 0.5$.

To further test how the results vary with the region selection, we extract the surface brightness profile
in small bin size ($6^{\prime\prime}$), and run the fit again. The compression factor remains intact
with $C = 2.1$. The profile is also extracted in a spherical sector instead of the elliptical one, although
the latter appears to better describe the morphology of the shock structure. The best-fit $C$ becomes 2.0, which is still consistent with the original value. The position of the edge also
remains the same. We further change the opening angle 
of the extraction sector from 70 degree to 50 degree or 90 degree, and find that both the compression
factor and edge position are consistent with the original value within errors. Therefore the 
obtained properties of the southeast shock are nearly independent on the detailed region selection.

To verify the surface brightness discontinuity found in the {\it Chandra} image, we extract the {\it XMM-Newton}
surface brightness profiles in the southeast and northwest cone regions (Supplementary Figure~2). The radial binning is set to 
10$^{\prime\prime}$, and the radial distance is calculated from the collision axis of the merging system. 
The observed profiles and the best-fit models are shown in Supplementary Figure~4, as well as the models
obtained with the {\it Chandra} profiles. For the northwest region, the {\it XMM-Newton} profile can be fit with the $\beta$ model. 
The best-fit parameters of the $\beta$ model agree well with the {\it Chandra} results. The southeast profile shows 
a discontinuity at $r=4^{\prime}-5^{\prime}$. By fitting the southeast profile with a broken power law model (Eq.~\ref{eq:bknpow} in the Supplementary Information), 
the position of the discontinuity $R_{\rm edge}$ is $4.1^{\prime} \pm 0.2^{\prime}$, and the compression factor $C = 2.0 \pm 0.4$,
indicating a shock front with $\mathcal{M} = 1.7 \pm 0.3$. These values are all consistent with the {\it Chandra} results within 
the uncertainties. In both southeast and northwest regions, the {\it XMM-Newton} profiles show surface brightness excess over 
the best-fit model at $r \sim 6^{\prime}$, which is not seen in the {\it Chandra} plot. The nature of the excess is unclear. Ignoring
the excess does not affect the fits. 

The above {\it XMM-Newton} analysis is based on the surface brightness profiles extracted in the $0.5-8.0$~keV. We have repeated the
surface brightness fits with the profiles extracted in the $0.5-2.0$~keV so as to enhance the source-to-background ratio. 
The best-fit compression factor and shock Mach number are consistent with the original values.

\subsection*{Shock X-ray temperature jump}

To rule out the presence of a cold front at the surface brightness discontinuity, and to obtain more constraints on the 
shock Mach number, we measure the ICM temperature profiles of the southeast equatorial region. First,
we extract {\it Chandra} spectra from the regions defined in Supplementary Figure~2. Each spectrum is fit by one absorbed {\it cie} component, with temperature, emission measure,
and redshift free to vary. The average metallicity is fixed to 0.3~solar as it cannot be constrained with the {\it Chandra} 
spectrum. The temperature profile is shown in Supplementary Figure~5. The best-fit temperatures for the pre- and post-shock regions are $T_{\rm pre} = 4.5 \pm 0.9$ keV and
$T_{\rm post} = 7.1 \pm 1.5$ keV, respectively. The resulting temperature jump $T_{\rm post}/T_{\rm pre}$ is $1.6 \pm 0.5$.
We can therefore confirm the presence of a shock by ruling out the possibility of a cold front at the surface brightness 
discontinuity. 

The standard Rankine-Hugoniot shock model \cite{landau1959} 
predicts the temperature jump condition,
\begin{equation}
\frac{T_{\rm post}}{T_{\rm pre}} = \frac{[2\gamma\mathcal{M}^2 - (\gamma - 1)][(\gamma - 1)\mathcal{M}^2 + 2]}{(\gamma + 1)^2\mathcal{M}^2}
= \frac{5\mathcal{M}^4 + 14\mathcal{M}^2 - 3}{16\mathcal{M}^2},
\end{equation}
where the right hand expression assumes $\gamma = 5/3$. Therefore, the Mach number can be obtained from the temperature
jump,
\begin{equation}
\mathcal{M} = \sqrt[]{\frac{\left(8\frac{T_{\rm post}}{T_{\rm pre}}-7\right) + \sqrt[]{\left(8\frac{T_{\rm post}}{T_{\rm pre}} - 7 \right)^2 + 15}}{5}},
\label{eq:mfromkt}
\end{equation}
and the error in the Mach number is 
\begin{equation}
\delta\mathcal{M} = \frac{4\mathcal{M}}{\sqrt[]{\left(8\frac{T_{\rm post}}{T_{\rm pre}}-7\right)^2 + 15}} \delta\left(\frac{T_{\rm post}}{T_{\rm pre}}\right).
\label{eq:merrfromkt}
\end{equation}
The best-fit Mach number for the southeast shock is $1.6 \pm 0.4$. It is well consistent with the Mach number estimated from the {\it Chandra}
surface brightness profile. We then prove the presence of a weak shock propagating towards outskirt along the equatorial direction of 
the early merger system.

The above temperature jump is obtained from the projected spectra. The post-shock temperature might be underestimated as the 
cooler gas at temperature $T_{\rm pre}$ is projected on the post-shock region. 
However, the current {\it Chandra} data do not allow a deprojected analysis, which requires 
spectra with high total counts. The deprojected analysis needs to be done with the {\it XMM-Newton} data. 

Next we measure the temperature jump at the southeast shock with the {\it XMM-Newton} data, accounting for the projection effect. The spectra are extracted from the regions shown in Supplementary Figure~2. 
The direct deprojection model presented in \cite{sanders2007} (\url{https://www-xray.ast.cam.ac.uk/papers/dsdeproj/})
is used to correct the projection effect
for the inner regions. For each region, the ICM component is modelled by one SPEX {\it cie} component,
absorbed by the Galactic neutral material.
The temperature, metallicity, redshift, and the emission measure of the thermal component are left free in the fits. As shown in Supplementary Figure~5, the best-fit temperatures are $5.2 \pm 0.9$ keV and $8.1 \pm 1.4$ keV for the pre- and post-shock regions, respectively. 
Both temperatures are consistent, within the statistical uncertainties, with the {\it Chandra} values based
on a projected analysis. Using Eqs.~\ref{eq:mfromkt} and \ref{eq:merrfromkt} in the Supplementary Information, the shock Mach number with the temperature
jump is $1.6 \pm 0.4$. This value is identical to the {\it Chandra} result, and in good agreement with the Mach numbers
based on the surface brightness discontinuity measured with both instruments.

We further investigate the temperature jump by using \textit{Suzaku}, which has low and stable background hence enables 
us to investigate the regions of low X-ray surface brightness. 
We extract {\it Suzaku} spectra from the post-shock region as shown in Supplementary Figure~2. To reduce the photon contamination 
from the bright post-shock region, the pre-shock region is moved 1 arcmin to the southeast. As shown in Supplementary Figure~5, the resulting temperature is $T_{\rm pre}=4.2 \pm1.0$ keV and $T_{\rm post}=6.9\pm1.1$ keV for the pre- and post-shock regions, respectively. With the observed temperature jump and Eqs.~\ref{eq:mfromkt} and \ref{eq:merrfromkt} in the Supplementary Information, the Mach number is estimated to be $1.7 \pm 0.4$.
These values are consistent with the results obtained with \textit{Chandra} and \textit{XMM-Newton}.

Although the surface brightness at the northwest of the merger axis does not show an apparent discontinuity (Supplementary Figures~3 and 4), given the
hot spot in X-ray temperature and the radio spectrum flattening (Figure~3), 
naturally we would expect the presence of a shock northwest on the equatorial plane. Here we also measure the temperature
profile for the northwest equatorial region. Spectra are extracted from the regions defined in Supplementary Figure~2. As shown in Supplementary Figure~5,
the projected {\it Chandra} analysis gives $T_{\rm post} = 6.9 \pm 1.0$~keV and $T_{\rm pre} = 4.2 \pm 0.7$~keV, the
temperature ratio is then $1.6 \pm 0.4$, and the shock Mach number can be estimated using Eqs.~\ref{eq:mfromkt} and \ref{eq:merrfromkt} in the Supplementary Information
as $\mathcal{M} = 1.6 \pm 0.3$. The {\it XMM-Newton} data allow a deprojection analysis. The best-fit temperatures with the
{\it XMM-Newton} spectra are $T_{\rm post} = 7.1 \pm 0.9$~keV and $T_{\rm pre} = 4.8 \pm 0.9$~keV, which give
a shock Mach number of $\mathcal{M} = 1.5 \pm 0.3$. We skip the {\it Suzaku} data as it does not cover the pre-shock region. 
Therefore, if the northwest shock exists, it would have nearly the same speed as the observed equatorial shock at southeast. 

\subsection*{Shock heating and the ICM thermal/ionization equilibrium}

Combining the {\it Chandra}, {\it XMM-Newton}, and {\it Suzaku} results for the southeast region, the mean
pre-shock temperature weighted by their errors is $4.6 \pm 0.7$ keV, and the mean post-shock temperature is 
$7.4 \pm 1.3$ keV. This temperature jump indicates a Mach number of $1.6 \pm 0.3$, and a compression factor
$C = 1.9 \pm 0.4$ at the shock front. The pressure jump at the shock is $3.1 \pm 0.9$, marginally consistent with
the observed value in the pseudo-pressure map (Figure~3). 
These are adopted as the best values. The measured 
emission measure indicates a mean electron density $n_{\rm e} = 2.0 \times 10^{-4} (l/2~\rm Mpc)^{-1/2}$ cm$^{-3}$
in the pre-shock region, where $l$ is the line-of-sight depth of the cluster volume. Assuming $l = 2$~Mpc,
and given the observed surface brightness jump at the shock front, the electron density in the post-shock 
region is $\sim 3.8 \times 10^{-4}$ cm$^{-3}$. As the shock has a low Mach number, the electrons are heated
by adiabatic compression,
\begin{equation}
n_{\rm e, post}^{1-\gamma} T_{\rm post} = n_{\rm e, pre}^{1-\gamma} T_{\rm pre},
\label{eq:adiaba}
\end{equation}
the adiabatic heating gives an expected $T_{\rm post}/T_{\rm pre} = 1.5$. Therefore, the observed temperature jump at the 
surface brightness discontinuity 
can be roughly explained by the adiabatic compression from the equatorial shock.

Converting the pre-shock temperature to the sound speed ($c_{\rm s} = 1088$ km s$^{-1}$), the observed shock velocity $v_{\rm s}$ is 
then $\sim 1740$ km s$^{-1}$. The velocity of the post-shock gas relative to the shock
is $v_{\rm p} = c_{\rm s} / C = 566$ km s$^{-1}$. Assuming that the shock propagates with a uniform speed from the collision axis, 
the age of the shock is then about 250 Myr. In reality, the propagation time is probably longer, 
because the speed of equatorial shock is expected to increase in time as the two subclusters approach \cite{ha2018}. Therefore, the current age estimate should be treated as a lower limit.

It is still uncertain how fast the equilibrium between the electron and ion temperatures is reached after 
the merger shocks. Assuming that the electrons are not heated effectively in the collisionless shock,  
given the low ICM density at cluster outskirt and high shock speed, it is possible that the post-shock region has not had enough
time to equilibrate via Coulomb collisions \cite{fox1997,takizawa1998,rudd2009,akahori2010,wong2009, wang2018}. 
The timescale
for collisional thermal equilibrium is  
\begin{equation}
t_{\rm eq} \approx 2 \times 10^{9} \left(\frac{T}{10^{8} \rm K}\right)^{3/2} \left(\frac{n_{\rm p}}{10^{-4} \rm cm^{-3}} \right)^{-1}
\left( \frac{\rm ln \Lambda}{40}\right)^{-1} {\rm yr},
\end{equation}
where $n_{\rm p}$ is the proton density ($n_{\rm p} \approx 0.83n_{\rm e}$), 
and $\rm ln \Lambda \approx 40$ is the 
Coulomb logarithm. Adapting the measured post-shock proton density and temperature, the time to reach thermal equilibrium is 
$t_{\rm eq} \sim 500$ Myr. Given the uncertainty in the shock age estimate, it is not clear 
if an overall electron-ion thermal equilibrium is reached in the post-shock region. 
Locally, the thickness of the possible non-thermal-equilibrium region can
be approximated as $v_{\rm p}t_{\rm eq} = 290$ kpc (or $\sim 2.9^{\prime}$ if the shock is propagate along
the plane of the sky).

Another possible non-equilibrium effect is the departure from collisional ionization equilibrium, as the 
charge state distribution of the ions might lag behind the electron temperature increase due to the low collisional
frequency at the shock. The equilibrium ionization timescale is $t_{\rm ie} = 3 \times 10^{9} (n_{\rm e}/10^{-4}$ $\rm cm^{-3})^{-1}$~yr.
Given the post-shock density $n_{\rm e, post} = 3.8 \times 10^{-4}$ cm$^{-3}$, it would take about 790~Myr for the ionization equilibrium.  

If exists, the non-ionization-equilibrium might affect the X-ray spectrum \cite{wong2009}. 
We refit the deprojected 
{\it XMM-Newton} spectrum of the post-shock region, and allow the {\it rt} parameter to vary during the fit.
$rt$ is the ratio between the ionization temperature and the electron temperature, $rt=1$ indicates an ionization
equilibrium. The best-fit $rt$ is $0.92 \pm 0.26$, suggesting that the post-shock ICM is not strongly
biased from equilibrium. A deep observation with future high-resolution X-ray spectrometers (e.g., XRISM, 
Athena) will be required 
to accurately determine the gas ionization state at the shock.

\subsection*{Re-accelerated radio tails associated with a member AGN}

\begin{figure*}[!htbp]
\begin{center}
\includegraphics[angle=0,width=\columnwidth]{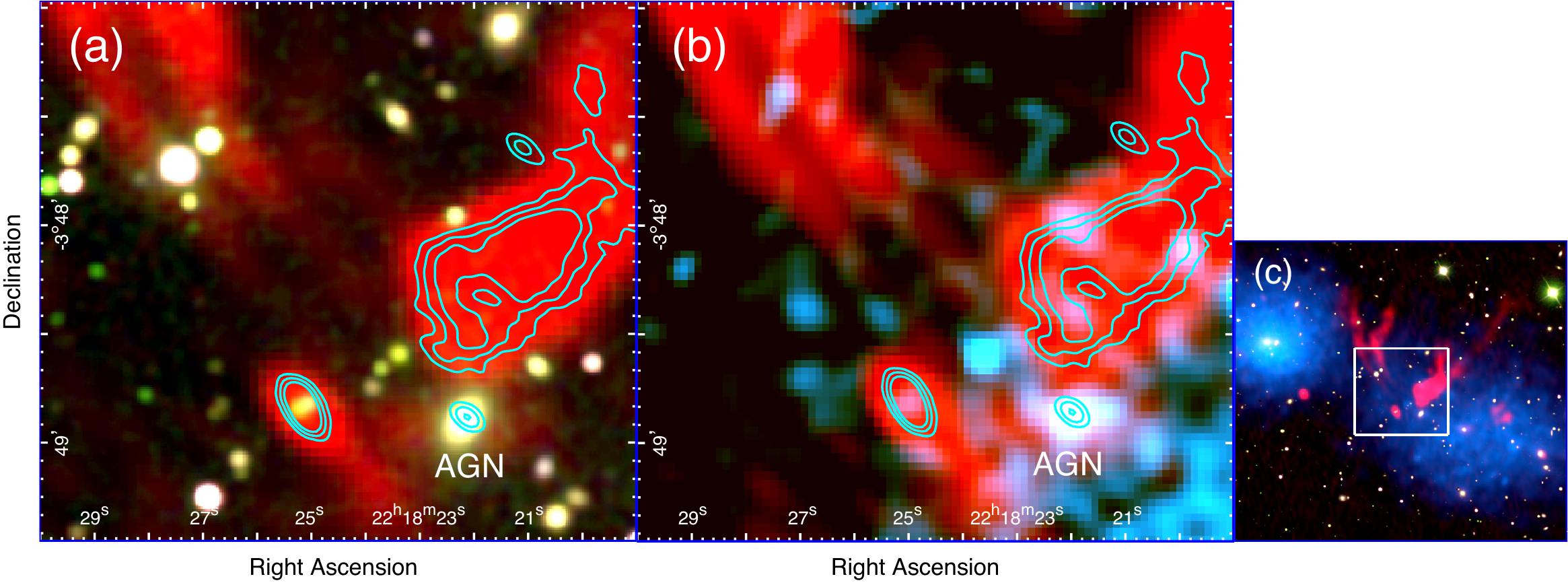}
\caption{ \textbf{Composite images of the region between 1E~2216.0-0401 and 1E~2215.7-0404.}
(a) Composite SDSS $gri$ image overlaid with the 235~MHz GMRT (red) and 610~MHz GMRT (contour) showing
the possible connection between the radio tails and the radio AGN. (b) {\it Chandra} image of the same field
overlaid with the 235~MHz GMRT (red) and 610~MHz GMRT (contour)
showing the diffuse gas halo associated with the radio galaxy.
The radio contours are given at the levels of $\sqrt{[1,8,64]}$ $\times$ 4$\sigma_{\rm rms}$, with $\sigma_{\rm rms}$ being the map noise.
(c) The white rectangle shows the field of view of panels (a) and (b).}
\label{fig:composite2}
\end{center}
\end{figure*}

The low-frequency radio sky of the pre-merger cluster is dominated by a pair of 330~kpc-long radio tails 
(source A in Figure~2), 
and a bright patch with a 450~kpc-long collimated narrow extension towards the northwest (source B).
As shown in Supplementary Figure~\ref{fig:composite2}, the radio tails are likely fueled by the host radio galaxy 
2MASX J22182217-0348539 at RA~=~22h18m22.19s, Dec~=~-03d48m54.0s. This galaxy also exhibits
a compact radio source at the 610~MHz. The optical spectrum of the host galaxy was taken in the 6dF galaxy survey \cite{jones2009},
which gives a spectroscopic redshift of $0.09434 \pm 0.00015$. Therefore the host galaxy
is likely a member of the merger system. The radio galaxy also hosts a small diffuse gas halo visible in
the X-ray. The {\it Chandra} image further shows a hint for a narrow gaseous extension that leads from the AGN
towards the northeast. Spectral analysis of the radio galaxy is not plausible with the current
X-ray data.


The LOFAR 144~MHz, GMRT 235~MHz and 325~MHz images are used to calculate the spectral index map. The radio images
at 610~MHz and higher are not deep enough to detect the majority of the steep spectrum emission. To correct for different sampling densities in the uv-plane, we use a uniform weighting,
and apply inner uv-range cuts to image only the common uv-range. The images are convolved to the lowest
resolution. A power-law model is used to fit the image pixels where all the three data are significant above
5$\sigma_{\rm rms}$. Any possible curvatures are ignored in the fits. 

The very steep ($-1.8 \sim -2.5$) spectrum of the radio tails indicate that the plasma has experienced 
significant synchrotron and inverse Compton losses. Naturally, one might expect that more aged plasma
would show a steeper radio spectrum. However, as seen in Supplementary Figure~\ref{fig:spidx}, there is no evidence for 
the expected spectral steepening towards the tips of the tails, where the older relativistic electrons
reside. The right tail of source A has a relatively
constant spectral index of $-1.8 \pm 0.1$, while the left one seems to be systematically steeper, and exhibits 
a patchy morphology in the spectral index map. It is hence difficult to explain the tails by radiative cooling alone. 
Provided that the radio tails further coincide, in projection, with the heated thermal structure in the ICM, we 
speculate that the present radio source is 
revived from faded electrons, by either a passing weak shock or an adiabatic compression caused by the merger.
The spectral index and morphology of source A seem to be generally consistent with a radio phoenix \cite{ensslin2001, ensslin2002, slee2001, vw2011, ogrean2011},
or some sort of old AGN tails re-energized through more gentle processes \cite{fdg2017}.

\begin{figure*}[!htbp]
\begin{center}
\includegraphics[angle=0,width=0.9\columnwidth]{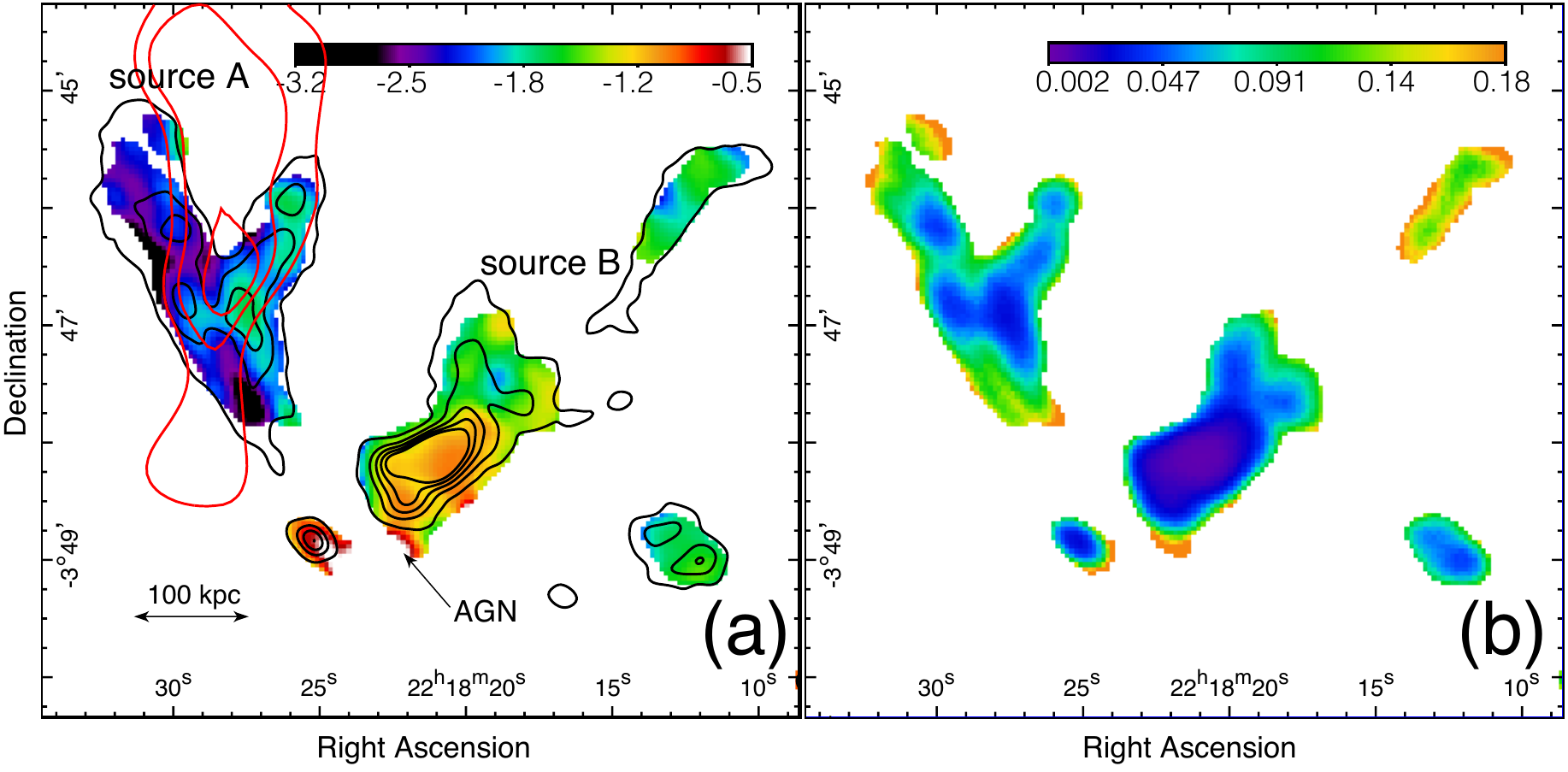}
\caption{\textbf{Radio spectral index map and spectral index uncertainty map of the radio sources.}
(a) Spectral index map at a resolution of 17.6$^{\prime\prime} \times 10.7^{\prime\prime}$ calculated between 144~MHz and 325~MHz. Black contours show the 325~MHz GMRT image (same as in Figure~2), and the red contours are the heated ICM structure
from the {\it XMM-Newton} temperature map (same as in the inset of Figure~3). Pixels with values below 5$\sigma_{\rm rms}$ are blanked. 
(b) Spectral index uncertainty (1$\sigma$) map of the same region.}
\label{fig:spidx}
\end{center}
\end{figure*}

As ICM heating and particle (re-)acceleration at source A might have the same origin, 
it would be interesting to compare the observed thermal energy excess with the 
energy that powered the relativistic electrons. Assuming that the heated ICM structure at source A has 
an ellipsoid shape in 3-D, the thermal energy excess is $\sim 1 \times 10^{59}$ erg.
Taking the shock speed estimated from the temperature jump in the northwest region (Supplementary Figure~\ref{fig:shockkt}), 
the heating time is $\sim 100$ Myr, and the heating rate is $\sim 3\times 10^{43}$ erg s$^{-1}$. 
The total energy spent in accelerating electrons can be approximated by  
the combined radio and inverse Compton radiation power.
The observed flux of source A at 144~MHz is 0.94 Jy, and the average radio spectral index is 
-2.3. Ignoring the possible curvature, the radio spectrum has the form of 
$S_{\rm \nu} = 0.94 $ Jy $ (\rm \nu / 144~MHz)^{-2.3} $. The bolometric luminosity 
is therefore $\sim 7 \times 10^{41}$ erg s$^{-1}$, including a typical inverse Compton component might
increase the luminosity to double that amount \cite{finoguenov2010}. 
The small radio-to-thermal-excess ratio implies that
little energy goes into accelerating electrons relative to heating the ICM.

The radio structure to the west, source B, gives rise to a number of questions.  The origin 
of the source is uncertain. Does it originate from the same AGN as the radio tails, or some
other objects? If it is indeed a part of the merger, how was the 450~kpc-long collimated extension
formed in such a dynamic system? Is the radio spectrum shaped by re-acceleration, which manages to keep
a relatively flat spectral index along the collimated structure? Answering these questions 
requires further deep radio and X-ray observations. We defer the discussion on the source B to
a follow-up paper, since it is not apparently related to the equatorial shocks.

\subsection*{Energetics of the equatorial shocks}

The rate of kinetic energy flow through the southeast shock front can
be estimated as
\begin{equation}
\Delta F_{\rm K} = \frac{1}{2} \rho_{\rm pre} v_{\rm s}^3 \left(1 - \frac{1}{C^2}\right),
\end{equation}
where $\rho_{\rm pre}$ is the pre-shock ICM mass density. Adopting $n_{\rm e} = 2.0 \times 10^{-4}$ cm$^{-3}$ in 
the pre-shock region, shock speed $v_{\rm s} = 1740$ km s$^{-1}$, and $C = 1.9$, the energy flux is
then about $6.5 \times 10^{-4}$ erg cm$^{-2}$ s$^{-1}$. The surface area of the shock depends 
on the assumption of its shape. As a simple approximation, we may treat the equatorial shock as a 
conical structure, with a radius $r_{\rm c} \sim 130$ kpc, and height $h_{\rm c} = 180$ kpc. The
shock surface area can be estimated as $\pi r_{\rm c} \sqrt[]{r_{\rm c}^2 + h_{\rm c}^2} \sim 0.1$ Mpc$^2$. 
With this size, the total rate of the shock kinetic energy is about $6.2 \times 10^{44}$ erg s$^{-1}$. 
Instead, if the shock has an ellipsoid shape, the surface area becomes $\sim 0.13$ Mpc$^2$, and the total kinetic energy
flow is then $\sim 8.1 \times 10^{44}$ erg s$^{-1}$. The shock power is about two times of the combined
radiative X-ray luminosity of the two clusters.

The thermal energy excess in the post-shock region can be estimated as $\Delta E = 
n_{\rm g, post} (kT_{\rm post} - kT_{\rm pre}) V_{\rm post}$, where $n_{\rm g, post}$ is the post-shock 
gas density and $V_{\rm post}$ is the volume of the post-shock region. Using the values determined 
from the shock jump condition, and assuming an ellipsoid shape of the post-shock region, the 
shock energy dissipated in the post-shock region is $\sim 2.6 \times 10^{59}$ erg. Assuming a shock
cross time of $h_{\rm c} / v_{\rm s} \approx 100$ Myr, the heating rate is then about $8.2 \times 10^{43}$ erg s$^{-1}$.
Therefore, the shock energy thermalized in the current post-shock region ($r \leq 500$ kpc from the collision axis) is only 
10\% of the total
kinetic energy. The main bulk of the shock energy is expected to be dissipated at a large radius. 


\end{document}